\documentclass{mn2e}
\usepackage{epsfig}
\usepackage{epsf}
\usepackage{graphicx}

\title[CSM Interaction Model for Eta Car's Eruption]{A Model for the
  19th Century Eruption of Eta Carinae: CSM Interaction Like a
  Scaled-Down Type~IIn Supernova}

\author[Smith]{Nathan Smith\thanks{Email: nathans@as.arizona.edu} \\
  Steward Observatory, University of Arizona, 933 North Cherry Avenue,
  Tucson, AZ 85721, USA}

\begin{document}
\date{Accepted 0000, Received 0000, in original form 0000}
\pagerange{\pageref{firstpage}--\pageref{lastpage}} \pubyear{2002}
\def\arcdeg{\degr}
\maketitle
\label{firstpage}

\begin{abstract}

  This paper proposes a simple model for the 19th century eruption of
  Eta Carinae that consists of two components: (1) a strong wind
  ($\dot{M}=0.33 \, M_{\odot}$ yr$^{-1}$; $v_{\infty}$=200 km
  s$^{-1}$), blowing for 30 years, followed by (2) a $\sim$10$^{50}$
  erg explosion (10 $M_{\odot}$; 750-1000 km s$^{-1}$) occuring in
  1844.  The ensuing collision between the fast ejecta and the dense
  circumstellar material (CSM) causes an increase in brightness
  observed at the end of 1844, followed by a sustained high-luminosity
  phase lasting for 10-15 years that provides a close match to the
  observed historical light curve.  The emergent luminosity is powered
  by converting kinetic energy to radiation through CSM interaction,
  analogous to the process occurring in more luminous Type IIn
  supernovae, except with $\sim$10 times lower explosion energy and at
  slower speeds (causing a longer duration and lower emergent
  luminosity).  We demonstrate that such an explosive event not only
  provides a natural explanation for the light curve evolution, but
  also accounts for a number of puzzling attributes of the highly
  scrutinized Homunculus, including: (1) rough equipartition of total
  radiated and kinetic energy in the event, (2) the double-shell
  structure of the Homunculus, with a thin massive outer shell
  (corresponding to the coasting cold dense shell) and a thicker inner
  layer (between the cold dense shell and the reverse shock), (3) the
  apparent single age and Hubble-like flow of the Homunculus resulting
  from the thin swept-up shell, (4) the complex mottled appearance of
  the polar lobes in {\it Hubble Space Telescope} images, arising
  naturally from Raleigh-Taylor or Vishniac instabilities at the
  contact discontinuity of the shock, (5) efficient and rapid dust
  formation, which has been observed in the post-shock zones of
  Type~IIn supernovae, and (6) the fast (3000--5000 km s$^{-1}$)
  material outside the Homunculus, arising from the acceleration of
  the forward shock upon exiting the dense CSM.  In principle, the
  bipolar shape has already been explained following earlier studies
  of interacting winds, except that here the requisite pre-existing
  ``torus'' may be provided by periastron collisions occuring around
  the same time, and the CSM interaction occurs over only 10 years,
  producing a thin shell with the resulting structures then frozen-in
  to a homologously expanding bipolar nebula.  This self-consistent
  picture has a number of implications for other eruptive transients,
  many of which may also be powered by CSM interaction.  A key
  remaining unknown is the ultimate source of the 10$^{50}$ ergs of
  energy required in the explosion.

\end{abstract}


\section{INTRODUCTION}

The chief reason why $\eta$ Carinae is an object of perpetual
astrophysical mystery is that the cause of its so-called ``Great
Eruption'' in the mid-19th century remains unexplained.  During this
event (see Davidson \& Humphreys 1997; Humphreys et al.\ 1999; Smith
\& Frew 2011), the star increased its bolometric luminosity and is
thought to have exceeded the classical Eddington limit by a factor of
roughly 5 as it briefly became the second brightest star in the night
sky, despite its distance of $\sim$2.3 kpc (Smith 2006). During this
event, the star also created the bipolar Homunculus nebula, which has
been studied in exquisite detail with the {\it Hubble Space Telescope}
({\it HST}) and numerous other observatories.  The total radiated
energy of the eruption ($\int L dt = 10^{49.3}$ ergs; Smith et al.\
2011; Humphreys et al.\ 1999) and the kinetic energy of the expanding
Homunculus nebula (about 10$^{49.7}$ ergs; Smith et al.\ 2003b)
released by this eruption rivaled that of a normal supernova (SN), but
the star survived the event --- and there are clues from its extensive
nebulosity that it may have done this multiple times in the past
(e.g., Walborn 1976; Smith \& Morse 2004).

This eruption has been adopted as the prototype for a class of stellar
outbursts, a number of which have now been identified in other
galaxies (see Van Dyk 2005; Van Dyk \& Matheson 2012; Smith et al.\
2011).  These extragalactic $\eta$ Car analogs masquerade as Type~IIn
supernovae (SNe~IIn), and are sometimes called ``supernova impostors''
due to their regular discovery in SN searches.  This type of outburst
may be a common -- although very brief -- rite of passage in the
evolution of very massive stars, and could dominate the total mass
lost by a massive star during its lifetime (Smith \& Owocki 2006;
Kochanek 2011).  Furthermore, the mass loss mechanism may be nearly
independent of metallicity, and therefore may offer a mode of mass
loss even for metal-poor massive stars in the early universe (Smith \&
Owocki 2006; van Marle et al.\ 2008).  Eruptive mass-loss akin to the
Great Eruption of $\eta$~Car is also thought to occur within a few
years to decades before some of the most luminous SNe~IIn known (e.g.,
Smith et al.\ 2007, 2010a; Smith \& McCray 2007; Ofek et al.\ 2007;
Woosley et al.\ 2007; Chevalier \& Irwin 2011).  The underlying
instability that causes these events, however, remains unproven and
presents a fundamental roadblock in our understanding of stellar
evolution for massive stars in general, and Population III stars in
particular.

Recent observations imply that most of the mass or energy ejection in
the 19th century outburst of $\eta$ Carinae occurred over a very short
time.  The most telling are the extremely thin and dense walls of the
Homunculus nebula (Smith 2006) and the small dispersion in ejection
dates derived from its expansion (Morse et al.\ 2001).  These imply
that the main ejection phase lasted only about 5 yr or less.  Combined
with the large amount of mass contained in the thin walls of the polar
lobes of the Homunculus (more than 10 $M_{\odot}$; Smith et al.\
2003), this would require a mass-loss rate during the eruption in
excess of a few $M_{\odot}$ yr$^{-1}$.  Such enormous mass flux is
well beyond the capability of a line-driven wind (Castor et al.\ 1975;
Aerts et al.\ 2004), and may even exceed the capability of a
super-Eddington continuum-driven wind alone (Owocki, Gayley, \& Shaviv
2004).  The ratio of mechanical to radiated energy is well over unity
(roughly a factor of 3; see above), implying that the 19th century
event was more like an explosion than a normal radiation-driven wind
(Smith et al.\ 2003).  Very high speeds observed in the surrounding
CSM outside the Homunculus also seem to imply an explosive component
in the event, which is harder to explain in any wind scenario alone
(Smith 2008).  Moreover, reanalysis of the historical light curve
shows brief brightening events associated with times of periastron
(Smith \& Frew 2011) and the recent discovery of light echoes from the
Great Eruption and associated spectra challenge the conventional
interpretation of a wind-driven event (Rest et al.\ 2012).


Following the hypothesis that the Great Eruption was powered by an
explosive event rather than a steady wind, this paper explores
potential consequences for understanding the historical light curve of
$\eta$ Car (Smith \& Frew 2011) and the formation of various
structures in the complex Homuculus nebula.  This paper treats the
event as an explosion to see if it provides a viable explanation for
the observed light curve.  In this context, such an explosion must be
non-terminal so that the star's core remains in-tact, as the star is
still observed to be luminous at the present time.


The most significant consequence of assuming an explosive origin for
the event is that radiation from the conversion of ejecta kinetic
energy into radiation can circumvent the paradox of exceeding the
classical Eddington limit for 20 years (much longer than the dynamical
time), because the radiating atmosphere is no longer required to be
hydrostatic.  It can also help explain how $\eta$ Car was able to
drastically change its radiative luminosity much faster than a thermal
timescale during the brief precursor brightening events in 1843 and
1838 (Smith \& Frew 2011; Smith 2011).

The model suggested below invokes an unknown physical mechanism to
cause an explosion that suddenly injected a large amount of kinetic
energy. Although there exists no firmly-established theoretical basis
for this, it is supported by a number of observational consequences.
As shown below, an explosion that overtakes a previously existing
dense wind can provide a natural explanation for both the emergent
luminosity of the event and its evolution with time, as well as
several detailed physical and structural properties of the Homunculus
nebula.  A simple CSM interaction model is presented in \S 2.  In \S 3
we then discuss how we can explain a number of otherwise very puzzling
observed features of the Homunculus nebula, if we borrow
well-established shock physics and observational precedents from the
class of SNe~IIn.  Finally, in \S 4 we discuss some implications for
other eruptive transients, and we conclude with a summary in \S5.

\section{A SIMPLE WORKING MODEL: A DENSE WIND FOLLOWED BY AN
  EXPLOSION}

\subsection{Motivation}

Two sets of different observations motivate the specific type of
explosion-powered model suggested below, which is very different from
the traditional picture of a super-Eddington wind usually discussed
for $\eta$ Car's eruption (e.g., Davidson 1987; Davidson \& Humphreys
1997; Owocki et al.\ 2004).

The first set of observations is that a number of different clues
point toward an explosive component associated with the eruption of
$\eta$ Car.  As noted above, several observations --- like the high
ratio of mechanical to radiated energy, the large mass and kinetic
energy of the nebula apparently ejected in a very short time, the
brief brightening events associated with periastron, and the extremely
fast (3000--5000 km s$^{-1}$) material seen in the ejecta around
$\eta$ Car --- all point toward some sort of a brief explosion that
was associated with the event.  A super-Eddington continuum-driven
wind could in principle supply the mass needed for the Homunculus over
a $\sim$15 yr time period (Owocki et al.\ 2004; van Marle, Owocki, \&
Shaviv 2008, 2009), but doing that and also producing high outflow
speeds seems problematic.  This is because one expects the effective
escape speed from the star to drop as it approaches the Eddington
limit (Owocki \& Gayley 1997; Owocki et al.\ 2004), and also because
at such extreme mass-loss rates around 1 $M_{\odot}$ yr$^{-1}$ or more
that are required for a wind alone, the wind is in the regime of
photon tiring\footnote{The term ``photon tiring'' can be alternatively
  expressed as adiabatic cooling of the gas and radiation field.},
where much of the radiated energy is used to accelerate the mass (see
Owocki et al.\ 2004; Owocki \& Gayley 1997; van Marle et al.\ 2008,
2009).  Thus, a relatively slow and heavy wind is generally expected,
in contradiction to the very fast speeds with homologous
expansion. Multi-dimensional effects may help simulations of
super-Eddington winds reach the physical paramters of the 600 km
s$^{-1}$ Homunculus, but probably not the much faster material outside
it.  Adding an explosive component can explain the higher velocities
that are observed, plus a number of other observations that we discuss
in detail later on.

The second set of motivating observations is that many observers have
noted that narrow H Balmer lines and other properties in spectra of
core-collapse SNe~IIn are very reminiscent of the spectrum of
$\eta$~Car.  Objects like some of the most luminous and energetic SNe
known (such as SN~2006gy), as well as stellar eruptive transients that
are 4-5 orders of magnitude less luminous, all have aspects of their
visual-wavelength spectra that appear similar to spectra of $\eta$
Car.  Many of these also exhibit evidence in the spectrum of some very
fast material, even in LBV eruptions (like SN~2009ip; Smith et al.\
2010b).  Historically, this similarity may have caused genuine
core-collapse SNe to be misclassified as $\eta$ Car analogs, as in the
case of SN~1961V (see Smith et al.\ 2011; Kochanek 2011), where early
observers like Zwicky (1964) noted its similarity to $\eta$~Car.  This
similarity in spectra includes recent reports of the spectrum of light
echoes from the Great Eruption of $\eta$ Car itself (Rest et al.\
2012).  With similar spectra produced over such a huge range of
luminosity, it seems likely that CSM interaction may play an important
role in more than just the most luminous SNe~IIn (where CSM
interaction is the only viable engine).

Moreover, the example of SNe~IIn demonstrates that explosion kinetic
energy which is converted to visual-wavelength radiation through CSM
interaction can significantly boost the luminosity and extend the
duration of the bright phases of the SN light curve.  Indeed, the
total radiated energy in the case of SN~2006gy was more than 10$^{51}$
ergs in visual light alone, showing that the conversion of kinetic
energy to visual light can be extremely efficient (Falk \& Arnett
1977; Smith et al.\ 2010a; van Marle et al.\ 2010; Chevalier \& Irwin
2011).  This same mechamism should work efficiently for lower energy
explosions as well.

\subsection{CSM interaction powering the Great Eruption?}

This section discusses a model for $\eta$ Car's eruption that is akin
to the standard model for SNe IIn, where an explosion produces a shock
wave that expands into dense circumstellar material (CSM).  A key
ingredient is the presence of very dense CSM ahead of the shock wave,
ejected by the star before the explosion.  We consider two different
potential origins for this pre-shock CSM in the sections to follow.

Dense CSM slows the shock, and the resulting high densities in the
post-shock region allow the shock to become radiative.  With high
densities and optical depths, thermal energy is radiated away
primarily as visual-wavelength continuum emission.  This loss of
energy removes pressure support behind the forward shock, leading to a
very thin, dense, and rapidly cooling shell at the contact
discontinuity (usually referred to as the ``cold dense shell'', or
CDS; see Chugai et al.\ 2004; Chugai \& Danziger 1994).  This CDS is
pushed by ejecta entering the reverse shock, and it expands into the
CSM at a speed $V_{CDS}$.  In this scenario, the maximum emergent
continuum luminosity from CSM interaction is given by

\begin{equation}
L_{CSM} \ = \ \frac{1}{2} \, \dot{M} \, \frac{V_{CDS}^3}{V_W} \ = \ \frac{1}{2} \, w \, V_{CDS}^3
\end{equation}

\noindent where $V_{CDS}$ is the outward expansion speed of the CDS,
$V_W$ is the speed of the pre-shock wind, $\dot{M}$ is the mass-loss
rate of the wind, and $w$ = $\dot{M}/V_W$ is the so-called wind
density parameter (see Chugai et al.\ 2004; Chugai \& Danziger 1994;
Smith et al.\ 2010).\footnote{Note that this scenario where radiation
  escapes efficiently is somewhat different from a more extreme case
  where the radiation diffusion time is comparable to the expansion
  timescale, changing the shape of the light curve and requiring
  trapped photons to do significant work (Smith \& McCray 2007;
  Chevalier \& Irwin 2011; Falk \& Arnett 1977).}

Can this general scenario of CSM interaction for SNe~IIn also be
applied to $\eta$~Car?  The luminosity of the extended bright phase of
the Great Eruption was of order 2.5$\times$10$^7$ $L_{\odot}$.  Most
of the mass in this CSM interaction ends up in the CDS, and so
$V_{CDS}$ should correspond to the expansion speeds observed for most
of the mass of the resulting Homunculus, which is about 600 km
s$^{-1}$ (Smith 2006).  The observed luminosity of the Great Eruption
would then require a wind density parameter of $w \simeq 10^{18}$ g
cm$^{-1}$.  This is similar to values of $w$ required for luminous
core-collapse SNe~IIn. As long as $\eta$ Car can supply pre-shock CSM
of this density, then a CSM-interaction model is feasible to explain
the Great Eruption luminosity.

In the CSM-interaction models explored below, we assume that the total
mass involved in the CSM interaction is 20 $M_{\odot}$, with 10
$M_{\odot}$ in the first mass ejection (pre-eruption wind or explosion
1), and 10 $M_{\odot}$ in the 1843/1844 explosion.  While these values
appear extreme, the choice is dictated by the observed mass of the
Homunculus.  Smith et al.\ (2003) measured a mass in the Homunculus of
at least 12.5 $M_{\odot}$, but noted that this was a likely lower
limit due to the assumptions involved in deriving the total gas mass
from observations of dust.  Smith \& Ferland (2007) derived a total
mass for the Homunculus of 15--35 $M_{\odot}$ from models to explain
the density of the molecular hydrogen gas (mostly independent of the
dust mass), and noted that values in the lower range of 15--20
$M_{\odot}$ were favored.  Gomez et al.\ (2010, 2006) derived a likely
upper limit to the mass of around 40 $M_{\odot}$ based on sub-mm
observations of cool dust (this is a probable upper limit because some
of this cool dust might be located outside the Homunculus).  We
therefore adopt a value of 20 $M_{\odot}$ for the total mass in the
Homunculus in these models.

\begin{figure*}\begin{center}
\includegraphics[width=5.1in]{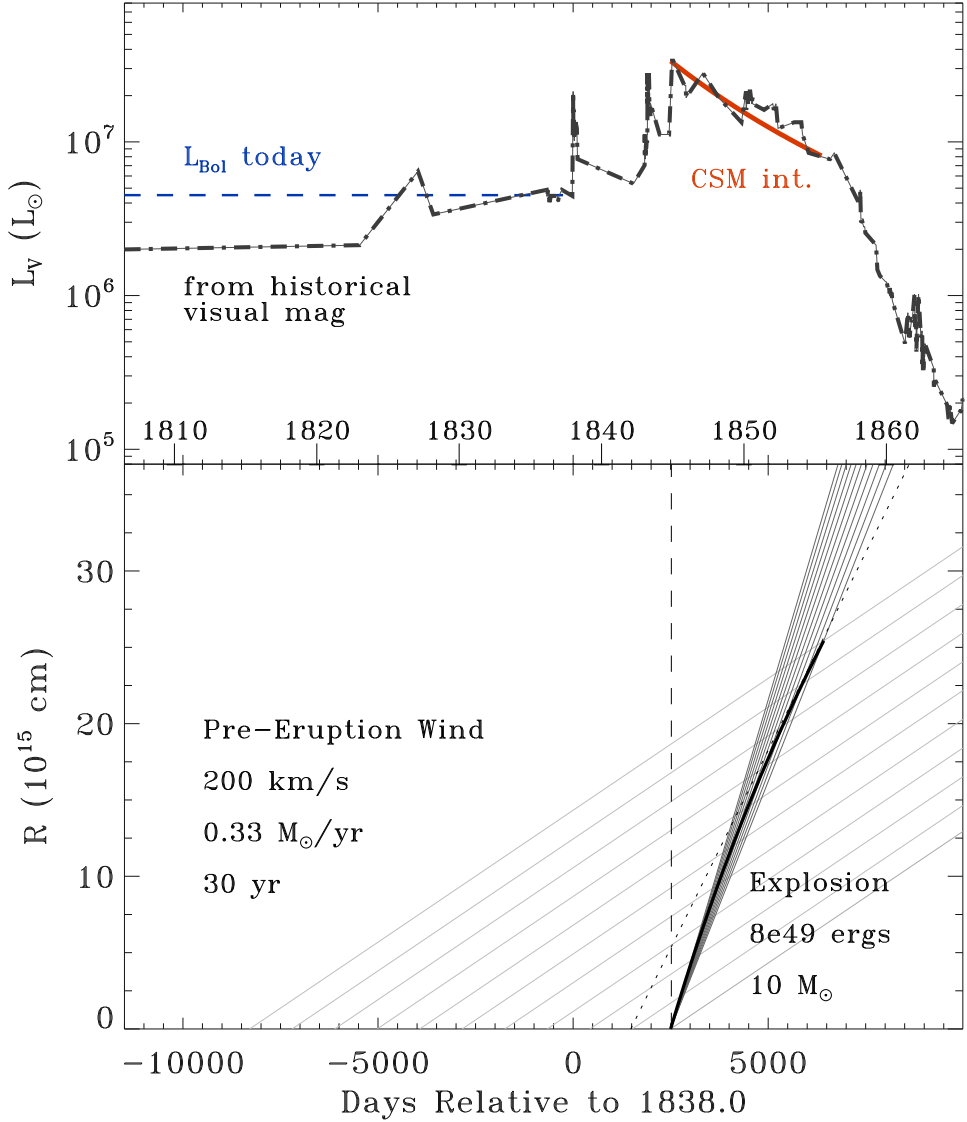}
\end{center}
\caption{A CSM-interaction model for the Great Eruption of $\eta$
  Carinae.  The top panel shows the historical visual-magnitude light
  curve of $\eta$ Car (Smith \& Frew 2011) converted from absolute
  visual magnitudes to solar luminosities.  This uses no bolometric
  correction, which is probably close to being valid for around
  1830-1860 (days $-$3000 to +8000 relative to 1838.0).  Before 1830
  the star was likely hotter than 7,000~K and so a bolometric
  correction is needed (the dashed blue line indicates its present-day
  bolometric luminosity), and after 1860 there is probably significant
  extinction from dust.  The orange curve shows the luminosity
  generated by CSM-interaction in our favored model, with an
  8$\times$10$^{49}$ erg explosion expanding into a dense wind.  The
  bottom panel shows the radius as a function of time, indicating
  trajectories for the two main components of the model: (1) a steady
  continuum-driven wind with $\dot{M} = 0.33 \, M_{\odot}$ yr$^{-1}$
  and $V_{\infty}$=200 km s$^{-1}$ (10 $M_{\odot}$ total) blowing for
  30 yr prior to 1844.0, and (2) an 8$\times$10$^{49}$ erg explosion
  in 1844, ejecting 10 $M_{\odot}$ at speeds ranging from 750 to 1000
  km s$^{-1}$.  The black solid curve is the resulting trajectory of
  the CDS, and the dotted black line marks the trajectory of the
  Homunculus that one would infer from today's observed polar
  expansion speed.  The trajectory of the CDS is calculated from
  conservation of momentum when the two components collide, and the
  resulting luminosity in the top panel is the difference in energy
  when momentum is conserved. }
\label{fig:model1}
\end{figure*}

\begin{figure*}\begin{center}
\includegraphics[width=5.1in]{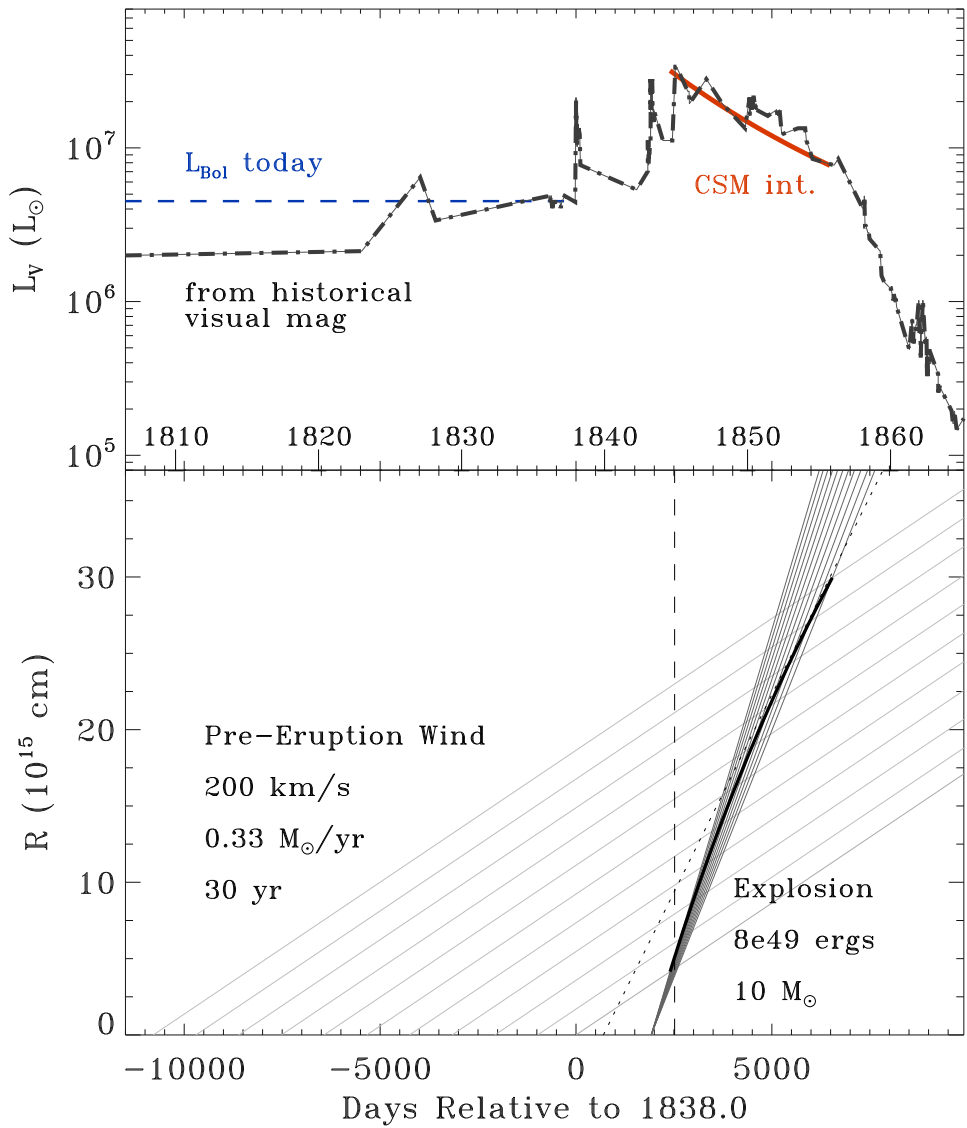}
\end{center}
\caption{Same as Figure~\ref{fig:model1}, but with slightly adjusted
  parameters in the times of ejection.  Here the main explosion
  coincides with the 1843 periastron passage instead of at the time of
  the 1844 brightening, whereas we assume that the dense wind shuts
  off in 1838 at the previous periastron event.  This is to provide a
  cavity so that the CSM interaction luminosity turns on after a
  delay, beginning in 1844.  The resulting light curve is identical to
  Model 1.  The purpose here is to illustrate that the model is not
  sensitive to the exact timing of the explosion (i.e. one can adjust
  the parameters slightly and still achieve a reasonable luminosity,
  providing that the CSM interaction begins at the onset of the
  brightening in 1844).}
\label{fig:model1b}
\end{figure*}

\subsection{Pre-shock CSM produced by a wind}

Let us first consider the simpler case where the required pre-shock
CSM is supplied by a very dense but steady wind, occurring for at
least a few decades before 1844.  In some sense, this is not far from
the traditional interpretation of LBV eruptions that invoke a
super-Eddington wind phase to drive the mass loss.  Here we assume
that this wind supplied much of the mass (about half), but very little
of the kinetic energy of the eruption.

A strong pre-eruption wind phase has observational support, since the
historical light curve shows that $\eta$~Car was already in a
long-duration eruptive state in the early 19th century (Smith \& Frew
2011).  The observed red-orange color of the star during the 1830s
($B-V = 0.7-1.2$ mag; Smith \& Frew 2011), with a known line-of-sight
reddening, suggests that the star had a cool temperature of
$\sim$7,000 K (and hence, a small bolometric correction), and that its
photospheric radius had swelled to be around 6$-$7 AU (Smith 2011).
This is 7--10 times larger than the present-day radius, and would
suggest a correspondingly smaller escape speed.  With a present-day
wind speed around 550 km s$^{-1}$ (Hillier et al.\ 2001), the
pre-eruption wind speed should then (naively) scale to about 160--220
km s$^{-1}$.  Based on the observed properties in the 1830s, we
therefore adopt a pre-eruption wind speed of order 200 km s$^{-1}$.
This is not a unique requirement of the observations, but it is
plausible.

It remains unclear if the emitting photospheric radius leading up to
the eruption is the hydrostatic radius of the star, or just a
pseudo-photosphere in the wind for which a drop in $V_{esc}$ is not
necessarily expected.  However, note that during the 1890 eruption,
when direct spectra of $\eta$~Car are available, the star exhibited a
cooler F-supergiant-like spectrum and -- more importantly -- Doppler
shifts of absorption lines seen in those spectra indicate an outflow
speed of only 200 km s$^{-1}$ (Whitney 1952; Walborn \& Liller 1977).
Humphreys et al.\ (1999) and others have commented that the properties
of the star during the 1890 eruption were probably similar to those at
the beginning of the Great Eruption, if one corrects for extinction
from the Homunculus in 1890.  Thus, there is a strong observed
precedent for the primary star having a slower and denser wind of 200
km s$^{-1}$ when it is in a cooler eruptive state.

\begin{table}\begin{center}\begin{minipage}{3.2in}
      \caption{Adopted input parameters for Models 1 and 2: Pre-shock
        CSM produced by a wind.}  \scriptsize
\begin{tabular}{@{}lccc}\hline\hline
Parameter   &Units          &Wind         &Explosion \\  \hline
Mass        &$M_{\odot}$     &10           &10        \\
$\dot{M}$   &$M_{\odot}$/yr  &0.33         &...       \\
$V_{exp}$    &km/s           &200          &750--1000 \\
$L_w$       &$L_{\odot}$     &1.1$\times$10$^6$ &... \\
$E_{kinetic}$&ergs           &4$\times$10$^{48}$ &7.7$\times$10$^{49}$ \\
\hline
\end{tabular}\label{tab:model1tab}
\end{minipage}\end{center}
\end{table}

For $V_W$=200 km s$^{-1}$, the corresponding mass-loss rate needed to
produce the fiducial wind density parameter of $w \simeq 10^{18}$ g
cm$^{-1}$ (see above) is roughly 0.3 $M_{\odot}$ yr$^{-1}$.  This
mass-loss rate is extremely high, and is much higher than can be
produced by normal line-driven winds (see Owocki et al.\ 2004; Smith
\& Owocki 2006; Aerts et al.\ 2004).  It is, however, comfortably
within the regime of continuum-driven winds that are mildy
super-Eddington ($\Gamma$ = 2 to 4; see Fig.\ 6 of Owocki et al.\
2004), and that experience some degree of photon tiring.  Such winds
have been discussed extensively in connection to $\eta$~Car and other
LBVs (Shaviv 2000; Owocki et al.\ 2004; Owocki \& Gayley 1997; van
Marle et al.\ 2008, 2009), as well as compact objects (Joss et al.\
1973; Quinn \& Paczynski 1985; Paczynski 1990; Belyanin 1999).  A
mass-loss rate of a few 10$^{-1}$ $M_{\odot}$ yr$^{-1}$ is plausible,
based on these studies of $\eta$ Car.  The photon-tiring limit (see
Owocki \& Gayley 1997), $\dot{M} = 2L/V_W$, is roughly 1.6 $M_{\odot}$
yr$^{-1}$ for $\eta$ Car's present luminosity (and $V_W$=200 km
s$^{-1}$), and would increase for higher values of $L$.  This adopted
pre-eruption wind used a luminosity $L_w = (1/2) \dot{M}$ (
$V_{\infty}^2+V_{esc}^2$) of roughly 2$\times$10$^6$ $L_{\odot}$.  If
the apparent temperature for the pre-eruption star was around 7,000~K,
as noted above, the bolometric correction was probably small, and so
the {\it emergent} radiation in the 1830s was likely to be close to
the star's present luminosity of $\sim$4.5$\times$10$^6$ $L_{\odot}$.
This means that the actual luminosity was at least 7$\times$10$^6$
$L_{\odot}$ and that at least 30\% of the available radiation energy
was used to accelerate the wind.

It is interesting to note that while slower wind speeds are one of the
obstacles to explaining the creation of the Homunculus nebula with a
wind acting alone, the slower speed is actually an advantage in the
CSM interaction model because a slower pre-shock CSM increases the
efficiency of converting shock kinetic energy into radiation.
Adopting $V_W$=200 km s$^{-1}$ and $\dot{M}$ = 0.3 $M_{\odot}$
yr$^{-1}$, we can then rewrite equation (1) as

\begin{equation}
  L_{CSM} = 2.5 \times 10^7 \ \big{(}\frac{\dot{M}}{0.3}\big{)} \big{(}\frac{V_{CDS}}{600}\big{)}^3 \ \big{(}\frac{V_W}{200}\big{)}^{-1} L_{\odot},
\end{equation}

\noindent with $\dot{M}$ expressed in $M_{\odot}$ yr$^{-1}$, and both
$V_{CDS}$ and $V_W$ in km s$^{-1}$.  Equation (2) shows that values of
roughly $\dot{M} \ = \ 0.3 \ M_{\odot}$ yr$^{-1}$ and $V_W$ = 200 km
s$^{-1}$ are plausible for a scenario where an explosion being driven
into a slow dense wind powers $\eta$ Car's Great Eruption luminosity.
We consider this hypothesis below in more detail.

Equations (1) and (2) are somewhat idealized, since the value of
$V_{CDS}$ will become slower with time as the shock sweeps up more of
the slow and dense CSM, and as the speed of explosion ejecta entering
the reverse shock slows as well (see, e.g., van Marle et al.\ 2010;
Smith et al.\ 2010a; Chugai et al.\ 2004; Chugai \& Danziger 1994).
Thus, instead of the light curve showing a flat 10 yr plateau at
2.5$\times$10$^7$ $L_{\odot}$, one might expect the luminosity to drop
slowly with time (as is indeed observed).

This effect is explored using the model shown in
Figure~\ref{fig:model1}, where we calculate the relevant physical
parameters at each time step.  This model has two phases, as presumed
above.  Phase 1 is a dense continuum-driven wind that is active for 30
years preceding the explosion, with parameters of $\dot{M}$ = 0.33
$M_{\odot}$ yr$^{-1}$ (chosen for convenience to yield a total mass of
10 $M_{\odot}$ in the CSM) and a terminal speed of 200 km s$^{-1}$.
One could adjust the wind speed, mass-loss rate, and duration of the
wind slightly to produce a similarly dense and extended CSM, as long
as it roughly obeys equation (2).  Phase 2 is a dynamical explosion
that occurs at the time of the main brightening in late 1844 (see
Smith \& Frew 2011), with a total kinetic energy of
7.7$\times$10$^{49}$ ergs, ejecting a total of 10 $M_{\odot}$ in a
Hubble-like flow with speeds ranging from 750 to 1000 km s$^{-1}$.
The radial trajectories of the wind and explosion ejecta are shown in
the bottom plot in Figure~\ref{fig:model1}.  The thick black curve
shows the resulting trajectory of the CDS where parcels of mass from
the two phases meet.  Momentum is conserved at each time step in the
collision, and the deceleration leads to a loss of kinetic energy.  We
assume that the difference between the initial and final kinetic
energy in this collision is radiated away, and the radiated energy
produces a corresponding emergent luminosity at each time step.  The
corresponding bolometric luminosity supplied by this collision is
shown by the solid (orange) curve labeled ``CSM int.'' in the top
panel of Figure~\ref{fig:model1}.  As shown in
Figure~\ref{fig:model1}, a simple CSM interaction model with the
parameters listed in Table~\ref{tab:model1tab} provides an excellent
match to the observed shape and magnitude of $\eta$ Car's light
curve.\footnote{Note that we do not include a contribution from the
  pre-eruption stellar luminosity in the radiation energy budget of
  the eruption, because it is not clear that the star's radiative
  output remains high when most of the star's envelope is explosively
  ejected (i.e. the energy budget of the star must do work to rebuild
  the envelope after ejection).}

The drop in luminosity at the end of this bright phase (around day
6000-7000 in Figure~\ref{fig:model1}, or the late 1850's) is not
included explicitly in the calculated model.  It can, however, be
explained naturally with CSM interaction, as the shock/CDS overtakes
the outer edge of the densest part of the slow continuum-driven wind
(this is why we assumed a 30 yr duration for the wind).  As the wind
density parameter $w$ drops, so does the CSM interaction luminosity.
After exiting the dense slow wind, the expanding shock will presumably
encounter a faster and lower density wind, corresponding to the star
in its more normal state before the eruption began.  Indeed, the
fainter visual magnitudes before 1800 (Smith \& Frew 2011) would
suggest that the star was hotter, and therefore more compact with a
higher escape speed (perhaps similar to the present-day wind).  Faster
CSM will lead to a lower CSM-interaction luminosity, as will a less
dense wind.  Alternatively, the luminosity may also drop because the
speed of ejecta entering the reverse shock will slow down at late
times and the shock loses power.  Similar drops in luminosity are
observed in SNe~IIn that are thought to be powered by CSM interaction,
such as SN~1994W (Chugai et al.\ 2004), SN~2009kn (Kankare et al.\
2012), and SN~2011ht (Mauerhan et al.\ 2012; Roming et al.\ 2012).
The formation of new dust grains may cause additional extinction, but
this is not required to explain the initial drop in luminosity in the
late 1850s, and observations are inadequate to constrain the influence
of extinction.  Dust formation in the context of CSM interation is
discussed below in \S 3.6.

The thin dotted line in Figure~\ref{fig:model1} shows the linear
trajectory of the polar regions of the Homunculus that one would infer
from present-day proper motions of the expanding nebula in images.
From assuming ballistic motion of the nebula, one would infer an ejection
date of 1842-1843 in this model, even though the explosion actually
occurred in late 1844.

\subsection{Pre-shock CSM made by a wind (variation)}

In Figure~\ref{fig:model1b} we show Model 2, which is almost identical
to Model 1 except that we have adjusted the times of ejection somewhat
to explore the observational consequences.  This model has the
explosion occuring at the periastron passage of 1843, instead of at
the time of the main brightening in late 1844.  This model may be
attractive if the periastron passage instantly triggers the explosion,
although we are not necessarily advocating this.  In that case, to
make the CSM interaction luminosity turn on in late 1844, as observed,
one would need to turn off the wind to allow a relatively low-density
cavity around the star.  Model 2 (Figure~\ref{fig:model1b}) shows that
if the dense wind is shut off at the time of the 1838 periastron
passage, the delay in turning on the CSM-interaction luminosity
matches the late 1844 brightening.  The trajectory of the CDS in this
model (dotted line in Figure~\ref{fig:model1b}) would lead a modern
observer to infer an ejection date at the end of 1839, assuming linear
motion.

The purpose here is not to advocate for this type of fine-tuning, but
to demonstrate that varying the exact time of the explosion and wind
properties by small amounts does not produce a major change in the
light curve (i.e. one can make adjustments and produce the same
result).  This is because the key parameters are the slow
pre-explosion wind speed, the high pre-explosion wind density, and the
energy of the explosion.  As we see below, significantly changing the
nature of the mass ejection that produces the CSM does introduce
significant changes to the light curve that are more difficult to
reconcile with observations.

\begin{figure*}\begin{center}
\includegraphics[width=5.1in]{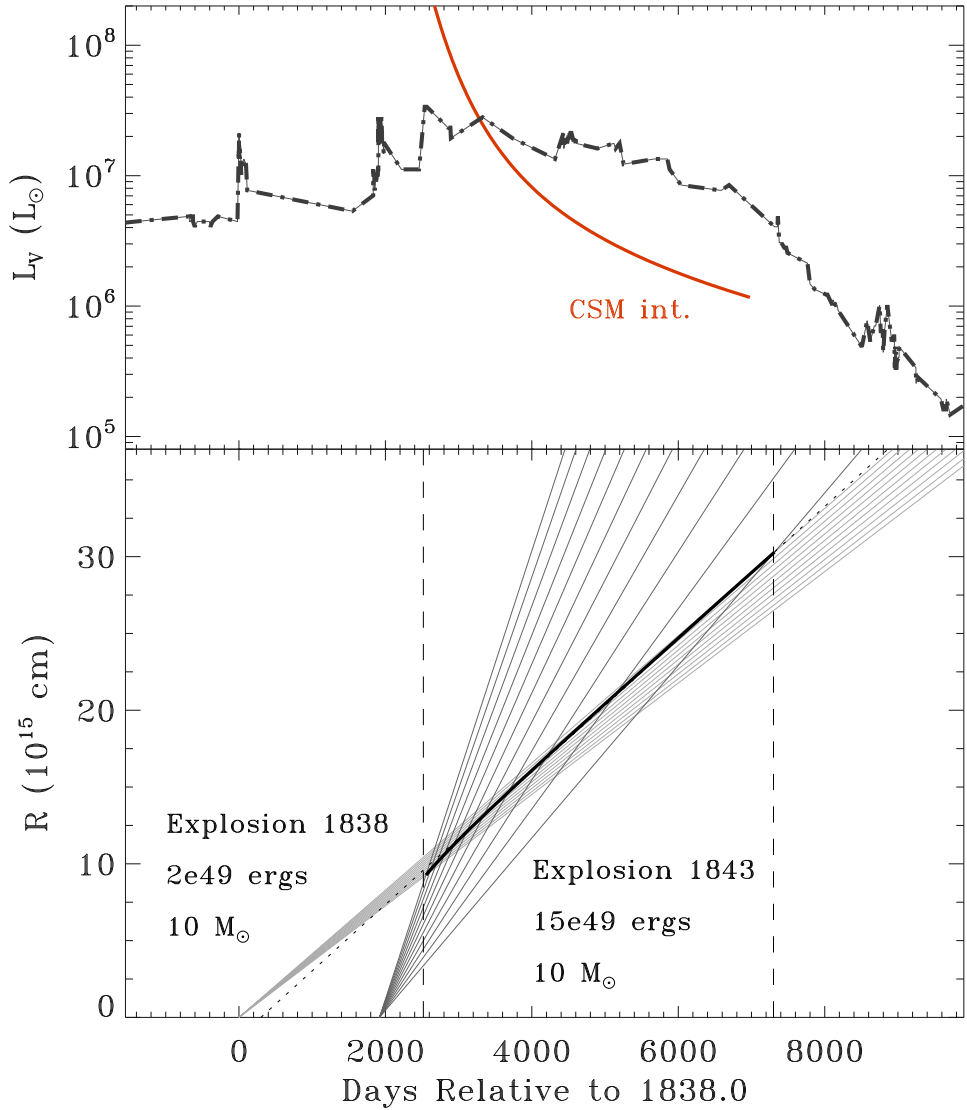}
\end{center}
\caption{Same as Figure~\ref{fig:model1}, but for an alternative model
  with two sequential explosive shell ejections that collide, instead
  of an explosion expanding into a dense wind.  This model produces a
  less satisfactory match to the emergent luminosity observed in the
  eruption.}
\label{fig:model2}
\end{figure*}

\subsection{Pre-shock CSM made by a previous explosion}

We also consider an alternative model (Model 3; Table 2) in
Figure~\ref{fig:model2}, where a sequence of two shells are ejected in
1838 and 1843 at periaston encounters when the secondary star plunges
into the envelope of the bloated primary star (see Smith 2011; Smith
\& Frew 2011).\footnote{Note that the two consecutive explosions of 10
  $M_{\odot}$ are adopted to calculate the resuting CSM interaction
  luminosity.  As noted by Smith (2011), the energy of the secondary
  star plunging through the primary star's envelope is insufficient to
  power the mass ejection, so some additional (and unknown) energy
  input would be needed.}  The collision of two shells in this way is,
in principle, similar to the model for SN~1994W suggested by Dessart
et al.\ (2009), except that SN~1994W was more luminous, shorter
duration, and a more energetic event.

The interaction and emergent luminosity were calculated in the same
way as above, but with the CSM provided by a previous explosion with a
range of speeds rather than a steady wind with all matter ejected at
the same speed.  Adopted parameters are listed in
Table~\ref{tab:model2tab}.  Figure~\ref{fig:model2} shows that a
scenario with two sequential explosive shell ejections does not
explain the observed light curve as well as an explosion crashing into
a dense wind.  It overproduces the luminosity at early times in the
event, and far underproduces the luminosity at late times.  This is
because the higher velocities involed drain more kinetic energy at
early times due to the $V_{CDS}^3$ dependence in Equation 1.  In other
words, at first the fastest ejecta from explosion 2 collide with the
slowest ejecta from explosion 1, producing a high luminosity due to
the large difference in speed --- whereas at later times, the slowest
ejecta from explosion 2 just barely overtake the fastest ejecta from
explosion 1, so the deceleration and mass involved in each time step
are much less severe, making it difficult to sustain a high luminosity
in this model.  

Moreover, this model requires significant and arbitrary fine-tuning in
the velocity ranges of the two explosions, in order to match the
observed duration of the bright event, unless some other
radiative-transfer effects are important.  The speeds that are needed
to make the CSM interaction have the correct duration also result in
smaller differences in speed, making the conversion of kinetic energy
to radiation less effecient; this in turn places higher demands on the
kinetic energy involved in the mass ejection (see
Table~\ref{tab:model2tab}), as compared to the wind + explosion model.
To make this double-explosion scenario produce the correct radiated
luminosity, one would need to carefully redistribute the mass in each
velocity range, and in a different way for both explosions
(i.e. relatively more mass ejected at high velocity in the 1838
explosion, and relatively more mass at lower speeds in the 1843
explosion). The trajectory of the CDS in this model (dotted line in
Figure~\ref{fig:model2}) would lead a modern observer to infer an
ejection date in 1839, assuming linear motion.

This double-explosion scenario therefore seems more difficult to
accomodate than the wind + explosion model presented in
Figure~\ref{fig:model1}, even though the sequential ejection of two
shells at two periastron events may provide a somewhat compelling
physical motivation.  The implication is that instead of dominating
the mass ejection and energetics, periastron passages may play an
important role in modifying the geometry (see \S 3.5).

\begin{table}\begin{center}\begin{minipage}{3.2in}
      \caption{Adopted input parameters for Model 3: Pre-shock CSM
        produced by a previous explosion.}  \scriptsize
\begin{tabular}{@{}lccc}\hline\hline
Parameter   &Units          &Wind         &Explosion \\  \hline
Mass        &$M_{\odot}$     &10           &10        \\
$V_{exp}$    &km/s           &420--480     &650--1700 \\
$E_{kinetic}$&ergs            &2$\times$10$^{49}$ &1.5$\times$10$^{50}$ \\
\hline
\end{tabular}\label{tab:model2tab}
\end{minipage}\end{center}
\end{table}

\subsection{Levels of complexity}

The favored model above (Model 1; Figure~\ref{fig:model1}) is quite
simple. One can, of course, fiddle with various parameters to adjust
the resulting light curve.  For example, the pre-eruption wind speed
$V_W$ may not be constant with time, or its mass-loss rate and wind
density may increase or decrease as the star brightens in the years
leading up to 1844 (see Smith \& Frew 2011).  Similarly, the mass in
the explosion may not have been distributed evenly across the range of
expansion speeds.  The goal here was to demonstrate that even the
simplest assumptions can provide a reasonable account of the high
luminosity and long duration of the Great Eruption.  This simple test
should be followed by more rigorous numerical hydrodynamical
simulations that more accurately account for radiative transport with
high optical depths, which may be important at early times.

In principle one could get a similar result by having a dense slow
wind that is followed by an ever increasing wind speed.  Such a
scenario would still produce a thin expanding shell that looked as
though it had a single age. This may apply to some extragalactic
LBV-like transients, since these are seen in large variety (see Smith
et al.\ 2011).  For example, the object UGC~2773-OT also has been
having an LBV-like eruption that is taking place over more than 10
years, although it has a much smoother light curve and no signs of a
fast blast wave in spectra (Smith et al.\ 2010b).  Ultimately, though,
this fast wind that follows the slow wind must carry a large amount of
kinetic energy in order to explain the visual light curve through CSM
interaction.  Most of the other extraglactic LBV-like transients have
shorter duration (less than 1 yr) and so steady winds are unlikely to
be suitable power sources.

Geometry adds another level of complexity, since one may account for
variations in mass and speed with latitude, as observations of the
Homunculus require (Smith 2006).  Soker (2007; and references therein)
has advocated a model wherein eruptive mass loss from the primary
caused a dense wind that is accreted onto the secondary star in a
binary system, in order to explain the Great Eruption of $\eta$ Car
and the formation of the Homunculus nebula.  In that model, the
radiative luminosity of the eruption is assumed to result from
accretion luminosity as 8 $M_{\odot}$ is accreted from the primary
wind onto the companion star through Bondi-Hoyle accretion over a few
years, and the bipolar shape of the Homunculus is caused by a fast
collimated wind or jet assumed to be launched by the secondary as a
result of that accretion.  A physical difficulty of this model is that
in order to explain the radiative luminosity of 2.5$\times$10$^7$
$L_{\odot}$ during the eruption, the $\sim$30 $M_{\odot}$ companion
star must accrete material at $\ga$20 times the Eddington accretion
rate, and accreting such a large amount of matter in such a short time
requires extreme parameters for the primary eruption that seem to
overwhelm any other energy source.  Some accretion onto the companion
may occur, and the impact on the geometry may be relevant, but it
seems impossible that accretion is the power source for the emitted
luminosity.  Instead, one can envision a version of the CSM
interaction model discussed above, where instead of an explosion, we
have a strong and fast wind acting over a short time, which overtakes
a slow dense wind emitted previously.  Making this into a collimated
fast wind or a jet is just a matter of the latitudinal structure of
the fast wind, and the arguments above would still hold.  The
different implications for the underlying physical cause of the
eruption are important, of course, but the end result of the CSM
interaction may be very similar.  Thus, it seems plausible that even
in a model that invokes a collimated fast wind or jet to help explain
the geometry, it is still quite likely that CSM interaction is the
engine that powers the radiated luminosity of the Great Eruption.

\section{OBSERVED CONSEQUENCES}

The Homunculus nebula around $\eta$ Car is one of the most intensively
observed objects in the sky (see the recent review by Smith 2009). Its
linear expansion extrapolates back to an ejection date during the
Great Eruption in the 1840s (Morse et al.\ 2001; Smith \& Gehrz 1998;
Currie et al.\ 1996), and as such, it provides some of our most
crucial information about the physics of the eruption.  The amount of
detailed information provided by the Homunculus is almost too rich:
While it is sometimes lamented that invoking binary systems allows
theorists to ``ascend into free parameter heaven'' (Gallagher 1989),
$\eta$~Car is a case where so much detailed information is available
that theorists may be weary of decending into an observationally
overconstrained purgatory that halts progress.  There are many
peculiar mysteries associated with the Homunculus, each presenting its
own challenge to one theory or another.

The approach below is to suggest that what works for traditional
SNe~IIn also seems to work well for $\eta$ Car and the Homunculus.
Essentially, we are suggesting that the Great Eruption of $\eta$~Car
behaved like a scaled-down (in kinetic and radiated energy) version of
the CSM interaction in a Type IIn supernova, and that the observed
properties of the Homunculus can be understood as the end result of
that CSM interaction.  The energy input comes from a weaker
non-terminal explosion of unspecified origin instead of Fe
core-collapse, but otherwise the shock physics is similar to a SN~IIn.
A number of well-established theoretical and physical precedents for
SNe~IIn can explain outstanding observational mysteries associated
with the Homunculus, as outlined below.

\subsection{Ratio of the kinetic energy of the Homunculus to the
  total radiated energy of the Great Eruption}

The first strong clue that the Great Eruption behaved more like an
explosion than a normal wind was its high observed ratio of kinetic
energy to radiated energy ($\zeta$=E$_{k}$/E$_{Rad}$), which is
substantially larger than unity (Smith et al.\ 2003).  It may be
possible to achieve $\zeta > 1$ with a continuum-driven wind that
suffers considerable photon tiring (Owocki et al.\ 2004; van Marle et
al.\ 2008, 2009), by using more than 2/3 of the available radiation
energy budget to accelerate the wind (i.e. including potential energy,
the luminosity required to accelerate the wind is $L = \dot{M}
(V_{\infty}^2 + V_{esc}^2)/2$), with the emergent radiation then being
considerably less than the total.  However, a ratio of $\zeta\ga 1$ is
also a natural consequence of strong CSM interaction, as seen in
SNe~IIn (Falk \& Arnett 1977; Smith \& McCray 2007; Chevalier \& Irwin
2011).  In a CSM-interaction model, the efficiency of converting
kinetic energy into radiation depends on the change in velocity as the
freely expanding ejecta ($V_{ej}$) cross the reverse shock and
decelerate to the speed of the CDS ($V_{CDS}$), as well as the
relative amounts of mass in the pre-shock CSM ($M_{CSM}$) and in the
explosion ejecta ($M_{ej}$).  For $M_{CSM} \ge M_{ej}$ and $V_{CSM} <<
V_{ej}$, the efficiency can aproach 100\%, for $M_{CSM} \approx
M_{ej}$ and $V_{CSM} < V_{ej}$, the efficiency can range from a few
per cent to roughly 50\%.  If $V_{CSM} \simeq V_{ej}$ then the
efficiency is very low, and of course, if $V_{CSM} \ge V_{ej}$ there
is no CSM interaction.  So, as long as the pre-1843 wind was dense and
slower than the explosion ejecta, CSM interaction is inevitable and an
efficiency of converting kinetic energy into radition of $\sim$25\% is
easily achieved.  Numerical simulations for SN IIn CSM interaction
have verified this (van Marle et al.\ 2010).

\begin{figure*}\begin{center}
\includegraphics[width=6.4in]{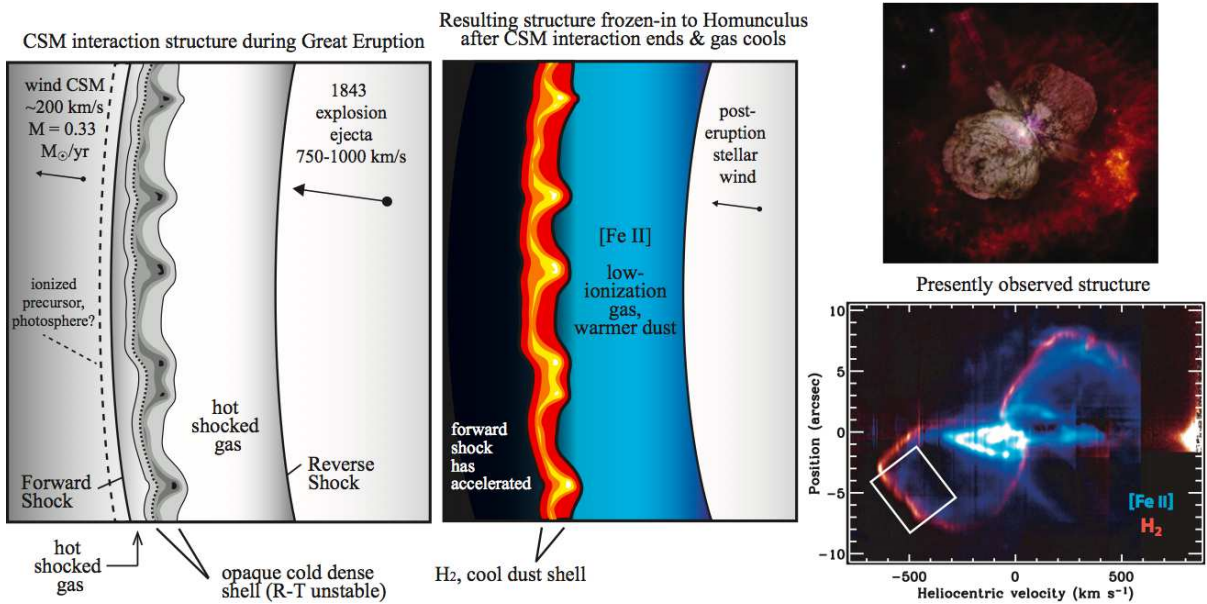}
\end{center}
\caption{Illustration of how CSM interaction determines the presently
  observed structure in the Homunculus.  The left panel shows a
  typical forward-shock/reverse-shock structure that arises in CSM
  interaction, as is often depicted for SNe~IIn (this is adapted from
  a sketch for SN~2006tf in Smith et al.\ 2008a).  A cold dense shell
  forms at the contact discontinuity between shocked CSM and shocked
  ejecta, which is Rayleigh-Taylor (RT) unstable (Chavalier \&
  Fransson 1994).  The middle panel shows the resulting density and
  ionization structure after the CSM interaction period ends.  The CDS
  contains most of the mass in a geometrically thin shell; it has
  cooled to form dust and molecules, and is optically thick to UV
  radiation so that dust remains relatively cool. The geometrically
  thicker inner shell corresponds to some of the reverse-shocked
  ejecta that has cooled to have neutral H but ionized metals, and the
  dust is warmer as it is exposed to the full luminosity of the
  central star.  The forward shock accelerated to higher speeds and
  reached much larger radii when it passed the outer boundary of the
  dense CSM shell.  The volume interior to the [Fe~{\sc ii}] shell is
  filled with post-eruption stellar wind.  The two panels on the right
  depict the presently observed structure of the Homunculus; on the
  top is an {\it HST} image showing the ``mottled'' structure of the
  polar lobes (see Morse et al.\ 1998; Smith et al.\ 2004), which we
  associate with RT instabilities in the CDS, and the lower panel
  shows a position-velocity plot of 2.122 $\mu$m H$_2$ and 1.644
  $\mu$m [Fe~{\sc ii}] emission observed with the Phoenix spectrograph
  on Gemini South (Smith 2006).  The white dashed box in the lower
  right highlights the double-shell structure that is sketched in the
  middle panel.}
\label{fig:sketch}
\end{figure*}

\subsection{Double-shell structure of the Homunculus}  

A clearly delineated double-shell structure exists in the Homunculus,
seen in tracers of dust (Smith et al.\ 2003) as well as in different
emission lines that trace different densities and ionization levels
(Smith 2006, 2002; Smith \& Ferland 2007).  Most of the mass (at least
12-15 $M_{\odot}$) resides in a geometrically thin outer shell
($\Delta R < 0.05 R$), whereas about 10\% of the mass resides in a
geometrically thicker ($\Delta R \simeq 0.2 R$) but optically thinner
inner shell.  The thin outer shell has cooler dust at $\sim$140~K,
whereas the inner shell has warmer grains that are closer to 200~K.
Also, the outer shell has neutral and molecular gas (depicted as
red-orange in Figure~\ref{fig:sketch}), whereas the inner shell has
neutral atomic H and singly ionized metals like Fe$^+$ (depicted as
blue in Figure~\ref{fig:sketch}). The thin outer shell and thicker
inner shell are also traced by UV absorption lines, with different
velocity widths and ionization levels in the two zones (Nielsen et
al.\ 2005; Gull et al.\ 2005).  Smith \& Ferland (2007) showed that
the observed ionization structure and dust-temperature stratification
in the walls of the Homunculus can be explained naturally by radiative
excitation (not shock excitation, as is usually assumed for gas
emitting near-IR H$_2$ and [Fe~{\sc ii}] emission), as long as the
outer shell has an abrupt increase in density compared to the inner
shell.  The origin of that density stratification, however, was not
explained.

We propose that even though the currently observed H$_2$ and [Fe~{\sc
  ii}] emission is not {\it powered} by shock excitation, the density
stratification that gives rise to this emission {\it is} the result of
shocks.  To clarify, we suggest that all the relevant CSM interaction
and shock heating occured in the 10-15 years after 1843.  The relevant
CSM-interaction shock structure during that event is depicted in the
left panel of Figure~\ref{fig:sketch}, which is the same as for a
generic SN~IIn (from Smith et al.\ 2008a; see also Chugai \& Danziger
1994; Chugai et al.\ 2004; Chevalier \& Irwin 2011; Chevalier \&
Fransson 1994).  After that time, the CSM interaction basically
stopped\footnote{Although see \S 3.7.} because the post eruption wind
($V_W \simeq 550$ km s$^{-1}$) had a speed comparable to or slower
than the CDS and its mass-loss rate was insignificant. Therefore, the
double-shell density structure corresponding to (1) the CDS, and (2)
to the thicker shell between the CDS and the reverse shock were
essentially ``frozen-in'' to the expanding structure.  The structure
is frozen in because the gas cools rapidly to 7,000~K and lower (dust
and molecules form), so the sound speed is much lower than the bulk
expansion speed of 600 km s$^{-1}$ and the flow is ballistic.  The
``frozen in'' density structure in coasting ejecta is a common feature
of simulations of spherical stellar explosions in a density gradient
(e.g., Fryxell, Mueller, \& Arnett 1991). The H$_2$ and [Fe~{\sc ii}]
emission structure we see today comes from coasting
radiatively-excited gas at different densities and with different
ionization/dissociation properties; shock excitation no longer plays a
relevant role in the energy balance.

\subsection{The apparent Hubble-like flow}

The main structural result of CSM-interaction with a radiative shock
is the formation of a dense and thin CDS, as noted above.  The
trajectory of the CDS in our model is shown with the thick black curve
in Figure~\ref{fig:model1}, and its expansion is nearly linear during
the eruption, with a final speed of about 530 km s$^{-1}$ in this
model.  When the most active phase of CSM interaction ends, we have
about 20 $M_{\odot}$ located in this fast-moving CDS.  Since CSM
interaction ends, this massive CDS will simply coast unless it sweeps
up a mass that is comparable, or until it encounters a high-pressure
region.  The CDS cooled quickly and will expand very supersonically
and ballistically for the 150 years after that, to be observed as the
Homunculus with the Hubble-like expansion we see today (Morse et al.\
2001; Smith \& Gehrz 1998; Currie et al.\ 1996).  The pre-eruption
wind and post-eruption wind have similar speeds and orders of
magnitude lower density, so they do not cause any perceptible
acceleration or deceleration of the shell after 1860. Thus, even
though the true mass loss was spread over a period of almost half a
century, the fact that it was swept-up into a thin shell through CSM
interaction (over a short time period that ended long before any
observations used to measure proper motions) makes it appear as if all
the mass in the Homunculus was ejected instantaneously.  In this
model, the ejection date one would measure is in 1842-1843, as noted
previously.  Thus, renewed investigations of the proper motion of the
Homunculus can provide an important test of when the explosion
actually occurred.

\subsection{Mottled structure of the polar lobes in high-resolution images}  

A very complex web of dark lanes and bright cells is seen on the face
of the south-east polar lobe of the Homunculus nebula, shown in the
upper right image in Figure~\ref{fig:sketch}.  The fine details of
this structure were first seen in refurbished {\it HST} images of
$\eta$ Car (Morse et al.\ 1998; although see also Duschl et al.\ 1995;
Currie et al.\ 1996; Smith et al.\ 2004).  Many admirers of $\eta$ Car
have since been perplexed by this complex network of structure,
resorting in some cases to vegetable comparisons or analogies to Solar
granulation in lieu of a physical mechanism.  This structure has no
obvious explanation in a wind-only origin for the Great Eruption.
While one does expect complex inhomogeneities in a super-Eddington
wind (Shaviv 2000; Owocki et al.\ 2004), one does not expect these
structures to persist at the same locations and size over 10 years
during the eruption.  Moreover, a wind-only mechanism would not give
rise to the sudden changes in outflow that are required for the
instabilities at a sharp interface like this.  On the other hand,
structures akin to this are a natural outcome of Raylieigh-Taylor
(R-T) instabilities that occur in a thin shell at the contact
discontinuity between a forward and reverse shock in CSM interaction
(e.g., Chevalier \& Fransson 1994).  Such structures --- including the
nonlinear thin-shell instability (Vishniac 1994) and the impulsive
case of R-T known as Richtmyer-Meshkov instabilities (Richtmyer 1960;
Meshkov 1969) --- are seen in a wide array of hydrodynamic simulations
of shock-ISM collisions and wind-wind interaction (i.e. planetary
nebulae; e.g., Frank \& Mellema 1994; Mellema \& Frank 1995; Fryxell
et al.\ 1991), including some of those performed for $\eta$ Car (e.g.,
Langer et al.\ 1999).

We propose that R-T instabilities (or Vishniac and Richtmyer-Meshkov
instabilities) gave rise to a complex network of cells and filaments
in the CDS with typical azimuthal size scales of a few per cent of the
radius (roughly comparable to the thickness of the CDS).  A key
difference in this scenario compared to previous interacting-winds
simulations of the Homunculus (Frank et al.\ 1995, 1998; Dwarkadas \&
Balick 1998; Langer et al.\ 1999) is that the shock is highly
radiative (as it must be to produce the emergent luminosity of the
Great Eruption), and the cooling causes material behind the forward
shock to collapse into an extremely thin cold dense shell (CDS). This
is the standard interpretation for SNe~IIn.  As a result, the
instabilities in the CDS have a much smaller radial extent than in
most interacting winds scenarios, more appropriate for the very thin
molecular layer of the Homunculus (Smith 2006).  An extremely thin CDS
corrugated by instabilities is seen in simulations of SN~IIn
collisions with significant cooling (van Marle et al.\ 2010).  In
$\eta$ Car, this complex structure was determined only during the
10-15 years of active CSM interaction during the Great Eruption; when
the CSM interaction stopped and the ejecta rapidly cooled, this
structure was again ``frozen-in'' to the ballistically expanding shell
for the next 150 years.  The post-eruption wind could have little
influence on even the small-scale structure of the polar lobes, since
the wind density dropped by more than three orders of magnitude and
had a similar expansion speed.  Along with rapid cooling came dust
formation in these R-T cells and filaments in the CDS, discussed
below.  That distribution of dust gives rise to the structure seen on
the surfaces of the polar lobes in {\it HST} images, since the
visual-wavelength light from the Homunculus is dominated by photons
scattered off dust grains.

Some observers have noted that the complex structure on the side walls
of the north-west polar lobe that we can see is different from that on
the face of the south-east polar lobe (e.g., Morse et al.\ 1998).
This difference may also be line with expectations of CSM interaction,
since the side walls of the polar lobes have oblique shocks that may
be dominated more by Kelvin-Helmholz (K-H) instabilities rather than
face-on R-T instabilities.  The transition between the two regimes
appears to occur at latitudes of roughly 45$\arcdeg$.

\subsection{Bipolar shape of the Homunculus}

In principle, the bipolar shape of the Homunculus has already been
explained theoretically in numerical simulations (Frank et al.\ 1995,
1998; Dwarkadas \& Balick 1998; Langer et al.\ 1999; Gonzalez et al.\
2004, 2010).  These simulations involved a scenario of a fast wind
blowing into a slow dense wind with an equatorial density enhancement
(or a bipolar fast wind blowing into a spherical slow wind in the case
of Frank et al.\ 1998).  While these simulations adequately reproduce
the overall bipolar {\it shape} of the Homunculus, there are some
problems reconciling them with detailed observations (see Smith 2006).
In general, the post-eruption wind of 1000-2000 km s$^{-1}$ (which is
faster than the observed post-eruption wind) overtakes a slower
pre-eruption wind, shaping the nebula over a period of $\sim$100-150
years.  The adopted mass-loss rates result in a Homunculus mass that
is an order of magnitude too small.  This has led to alternative
suggestions of an intrinsically bipolar wind during the Great Eruption
with a much higher mass-loss rate, where the bipolar shape and
latitudinal mass distribution result from the latitudinal variation of
escape speed and mass-loss rate on a rotating star (Dwarkadas \&
Owocki 2002; Owocki et al.\ 1996, 1998; Owocki \& Gayley 1997; Owocki
2003), which in principle also matches the observed shape (Smith 2002,
2006).

The scenario advocated here has some overlap with aspects of
interacting-winds simulations (especially with Frank et al.\ 1995),
except that instead of a post-eruption wind to inflate the Homunculus
acting over a century, we adopt an explosion that impulsively
accelerates the dense CSM/wind, and the mass and kinetic energy
involved are much higher.  This causes a key difference between our
proposed scenario and these interacting winds models, which is that
{\it the shaping of the nebula occurs over only 10-15 yr}. The
collision conserves momentum, and the loss of energy to radiation over
this short time period gives rise to the extra luminosity of the Great
Eruption, which is not included in any of the previous
interacting-winds models.  After about 1860, the central star
presumably returns to its normal (weaker, present-day) wind and has
little influence, in contrast to the interacting winds models.

What about the equatorial density enhancement required in these
models?  Observations of the Homunculus show that most of the mass is
located over the poles, not in the equator (Smith 2006), and proper
motions show that the extended equatorial skirt is not an older
feature that was pre-existing around the Homunculus.  This is
problematic for models that create the bipolar shape with a fast wind
continuously sweeping into an older and extended disk-like wind.  In
simulations by Dwarkadas \& Balick (1998), however, a very small
(10$^{14}$ cm) torus surrounded the star before the eruption; this was
sufficient to collimate and deflect mass into a bipolar flow, and the
torus was destroyed in the interaction. Acting on such small size
scales (where photon trapping will be important), this is similar to
having an intrinsically bipolar explosion, at least as far as larger
size scales and coasting trajectories are concerned.  Perhaps a very
compact slow torus could have been created during repeated periastron
passages in 1838 and 1843, because during these periastron events the
secondary star must have plunged into the bloated photosphere of the
primary (Smith 2011).  Brief 100-d peaks in luminosity were observed
at times of periastron before the Great Eruption (Smith \& Frew 2011).
Simulations of this stellar collision would be very interesting, in
order to determine the expected geometry.  From studies of the central
binary system, it has been established that the orbital plane of the
binary is in fact the same as the equatorial plane of the Homunculus
(Madura et al.\ 2012).  The explosion geometry resulting from the
internal distribution of angular momentum in the star's envelope (see
Balbus \& Schaan 2012) and the time-dependent tidal influence of the
nearby companion star are essentially unexplored in this context.

The extended ``equatorial skirt'' is prominent in {\it HST} images of
the Homunculus (see Figure~\ref{fig:sketch}), and this structure is
almost unique to $\eta$ Car --- generally equatorial features like
this are not seen in planetary nebulae.  Some ideas have been proposed
to explain the creation of a flat equatorial skirt concurrently with
the lobes in a super-Eddington eruption (Smith \& Townsend 2007;
Shacham \& Shaviv 2012).  However, we also point out that the
equatorial skirt seen in {\it HST} images is somewhat illusory.
Mid-IR images of thermal emission and some emission-line spectroscopy
reveal that the equatorial skirt actually contains very little mass
(Smith et al.\ 1998, 2002, 2003; Polomski et al.\ 1999; Smith 2002,
2003, 2008; Zethson et al.\ 1999; Hartman et al.\ 2004), and may
result more from preferential illumination than from an equatorial
density enhancement.  Light escaping through a clumpy equatorial torus
at the pinched waist of the Homunculus may explain the radial streaks
with escaping beams of starlight.  Thus, the existence of the
equatorial skirt does not place important constraints on the geometry
of CSM interaction.

\subsection{Rapid and efficient dust formation}  

Massive dust shells are a commonly observed property of LBVs, and the
dusty Homunculus of $\eta$ Car is the best studied.  Kochanek (2011)
has discussed the efficient formation of large dust grains in an
eruptive LBV wind, requiring very high mass-loss rates above
10$^{-2.5}$ $M_{\odot}$ yr$^{-1}$.  In general, the formation of dust
in a constant-velocity wind places strong demands on the density and
mass-loss rate, which for hot stars can only be accomplished in the
super-Eddington winds envisioned for LBV eruptions.

However, there is another way to trigger efficient and rapid dust
formation that involves explosive mass loss rather than winds, and it
has a well-established observational precedent.  Namely, SNe with
strong CSM interaction are observed to rapidly form copious amounts of
dust in their post-shock layers. The first well-established case was
SN~2006jc, which simultaneously (beginning only 50 days after
explosion) showed IR excess from newly formed hot dust, increased
fading in its visible light curve, and the characteristic blueshift of
its narrow emission lines formed in CSM interaction (Smith et al.\
2008b).  SN~2006jc was a peculiar Type Ibn explosion (weak H lines),
but a number of SNe IIn have shown the same post-shock dust formation
(Smith et al.\ 2009, 2008a, 2012; Pozzo et al.\ 2004; Fox et al.\
2009, 2011; Mauerhan \& Smith 2012).  The {\it formation} of dust in
shocks is perhaps somewhat surprising, since one normally expects fast
SN shock waves to destroy dust.  The key difference in SNe~IIn is that
the CSM is very dense and the shock is slower and radiative.  The
efficient radiation from the shock also provides efficient cooling.
In turn, the cooling removes pressure support and the forward shock
collapses to a CDS, with densities above 10$^{10}$ cm$^{-3}$.  The
higher densities from shock compression and lower temperatures from
enhanced cooling of the radiative shock then lead to efficient dust
formation.

Particle densities of order 10$^{10}$ cm$^{-3}$ are required for
nucleation of dust grains (e.g., Clayton 1979), a condition which is
met in the cold dense shells of SNe~IIn that are observed to form
dust.  In the free expansion that has dominated the Homunculus in the
150 years since the end of the Great Eruption, the density in the
Homunculus should drop as the radius increases, scaling roughly as
$\rho \propto r^{-3}$.  Since the ejecta are in linear expansion
during this time, we can also expect $\rho \propto t^{-3}$.  At the
present epoch, densities of $\sim$10$^7$ cm$^{-3}$ are required in the
thin outer shell of the Homunculus in order for H$_2$ to survive
(Smith \& Ferland 2007).  Scaling from the present epoch $\sim$160
years after the initial explosion in 1843 to just 10 years after (when
CSM interaction ended), the density in the thin outer shell must have
been 16$^3$ times larger, or a few $\times$ 10$^{10}$ cm$^{3}$.  These
densities are sufficient for dust nucleation.  The gas temperature
must also drop below $\sim$1500 K, but this is easily achieved as the
luminosity drops in the late 1850s and early 1860s.

The resulting physical parameters in the cooled, dense post-shock gas
in CSM interaction satisfy the same conditions of temperature and
density as very dense LBV winds that are thought to rapidly form large
dust grains.  The CSM-interaction mode of post-shock dust formation
has a clear advantage over the constant-velocity wind hypothesis
(Kochanek 2011), however, which is that in CSM interaction, the dust
is expected to reside in a very thin shell corresponding to the CDS,
as observed for $\eta$ Car (Smith et al.\ 2003) and many other LBVs.
This may be a significant obstacle for the idea of dust formation in a
long-duration ($\sim$100 yr) wind proposed recently by Kochanek (2012)
to explain the 20th century light curve of $\eta$~Car.

Efficient dust formation in shocks has other precedents besides
SNe~IIn as well.  In particular, the colliding winds of Wolf-Rayet
binaries with WC stars form dust very effeciently due to the
compression of the shock (see Crowther 2007).  In fact, $\eta$ Car
itself provides an observed example of this phenomenon in its
present-day state, since it shows episodes of enhanced dust formation
in the colliding wind shock of the binary system during periastron
passages (Smith 2010).

\begin{figure}\begin{center}
\includegraphics[width=3.0in]{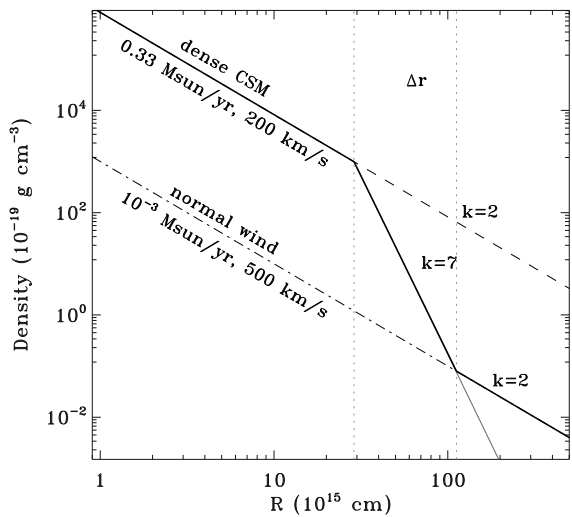}
\end{center}
\caption{Density as a function of radius.  The dashed line and
  dot-dashed line represent densities for two different mass-loss
  rates with $\rho \propto r^{-k}$ and $k$=2 (i.e. steady winds).  One
  represents the strong continuum-driven wind leading up to the
  eruption in Models 1 and 2, with $\dot{M}$=0.33 $M_{\odot}$
  yr$^{-1}$ and $V_W$=200 km s$^{-1}$, and the other is for the wind
  of $\eta$~Car in its normal (pre- and post-eruption) state, with
  $\dot{M}$=10$^{-3}$ $M_{\odot}$ yr$^{-1}$ and $V_W$=500 km s$^{-1}$.
  The thick black line shows a possible example of the transition
  between the two (see text), which has a much steeper density
  gradient with $k$=7 if we assume that the smooth transition took an
  additional 30 yr.  This is the pre-shock density at about 10-15 yr
  after the explosion, appropriate for the time when the forward shock
  reaches the outer extent of the dense CSM.  The pre-shock density
  drops by a factor of $\sim$10$^4$ over a range of $\Delta r$ in
  radius ($\Delta r$ depends on how quickly the dense wind turned on
  leading up to the eruption, but is unknown).}
\label{fig:density}
\end{figure}

\subsection{Some very fast material in a blast wave outside the
  Homunculus}

The perplexing structure in $\eta$~Car's ejecta continues outside the
Homunculus.  The so-called ``outer ejecta'' are a complex network of
ionized, N-enriched condensations that appear to have been ejected
centuries or millenia before the 19th century Great Eruption (Walborn
1976; Walborn et al.\ 1978; Walborn \& Blanco 1988; Davidson et al.\
1982, 1986; Meaburn et al.\ 1996; Weis 2001; Smith \& Morse 2004; Weis
et al.\ 1999, 2004; Smith et al.\ 2005).  The dense condensations seen
in images are expanding away from the star with typical speeds of a
few 10$^2$ km s$^{-1}$.

There is, however, also evidence for much faster material that was
ejected more recently, in the form of the long ``whiskers'' or
``strings'' seen in {\it HST} images (Morse et al.\ 1998; Weis et al.\
1999, 2004), which are expanding radially with speeds of $\sim$1000 km
s$^{-1}$, as well as the fastest material moving at 3000-5000 km
s$^{-1}$ (Smith 2008) that is seen only in long slit spectra because
it is Doppler shifted out of the narrow {\it HST} imaging filters.
This fast material resides outside the Homunculus but inside the dense
``outer ejecta'' condensations mentioned above. Smith (2008) suggested
that this fast material must arise from a blast wave from the Great
Eruption, and that the collision between this very fast material and
the slower and older condensations in the outer ejecta gives rise to
the soft X-ray shell seen around $\eta$~Car (Seward et al.\ 2001;
Corcoran et al.\ 1995, 2004).


In the CSM-interaction interpretation advocated here, this outer soft
X-ray shell would represent the current location of the forward shock.
The reason that the radius of the forward shock is so much larger than
the radius of the Homunculus now is that the forward shock accelerated
when it encountered a steep density gradient at the outer boundary of
the dense CSM, leaving the ballistically expanding CDS (and hence the
Homunculus) behind.  This situation is analogous to a rarefaction wave
that causes the acceleration of a SN blast wave when it passes through
and exits the dense envelope of a star (e.g., Matzner \& McKee
1999). In general, a shock front will accelerate when it encounters a
steeply falling density gradient (Gandel'man \& Frank 1956; Sedov
1959; Sakurai 1960; Colgate \& Johnson 1960; Imshennik \& Nad\"ezhin
1989; Ostriker \& McKee 1988; Chevalier 1992).  For a blast wave
propagating through a medium with a density gradient given by $\rho
\propto R^{-k}$, the velocity of a blast wave is described by

\begin{equation}
V_{BW} \propto R^{\frac{-(3-k)}{2}}
\end{equation}

\noindent (see Ostriker \& McKee 1988).  For steady winds with $k$=2,
the shock speed can be roughly constant or decelerate slowly, but for
steeper density power laws of $k > 3$ the shock front will accelerate.
Figure~\ref{fig:density} shows a plausible density structure for the
CSM around $\eta$ Car, appropriate for our first model of an explosion
running into a strong wind.  Figure~\ref{fig:density} shows a density
that switches from the wind of $\eta$ Car in its normal quiescent
state before the eruption (presumably similar to its present day
values of $\dot{M}$=10$^{-3}$ $M_{\odot}$ yr$^{-1}$, $V_W$ = 500 km
s$^{-1}$) to that of the assumed pre-eruption wind that produced the
dense CSM needed for our favored model of CSM interaction
($\dot{M}$=0.33 $M_{\odot}$ yr$^{-1}$, $V_W$ = 200 km s$^{-1}$).  If
we assume a steady transition in $\dot{M}$ that took place over a time
period of an additional 30 yr preceeding the dense wind, we get a
slope of $k$=7 (shown in Figure~\ref{fig:density}).  The density
gradient could have been much steeper than this if the beginnig of the
eruption was more abrupt.  Note that the drop in pre-shock density at
this outer radius of the CSM is also a necessity for explaining the
drop in luminosity at the end of the bright phase of the Great
Eruption, and a steeper slope than shown in Figure~\ref{fig:density}
would be commensurate with the observed fading (but of course, there
is also the possibility of dust formation or escape of radiation at
shorter wavelengths as the material becomes more optically thin).  The
resulting change in density is huge - roughly a factor of 10$^4$ over
the change in radius $\Delta r$ in Figure~\ref{fig:density}.  For this
conservative density power law of $k$=7, the scaling in equation (3)
indicates that the blast wave would accelerate from 550 km s$^{-1}$ at
$R \simeq$3$\times$10$^{16}$ cm to more than 8,000 km s$^{-1}$ at $R
\simeq$3$\times$10$^{16}$ cm.  For a steeper drop in density than this
hypothetical example, the blast wave acceleration would be more
severe. 

The resulting very fast ejecta arise from rapid expansion of a
relatively low mass ($\sim$0.1 $M_{\odot}$; Smith 2008) of very hot
gas located between the forward shock and the CDS, which supplies the
pressure of the blast wave at the moment when it reaches the outer
boundary of the dense CSM.  During the most intense phase of CSM
interaction (during the 10-15 yr after 1843), the forward shock would
have been located immediately ahead of the CDS, as depicted in the
left panel of Figure~\ref{fig:sketch}.  The gas between the forward
shock and the CDS is hot (10$^6$-10$^7$ K), but this zone is very
thin.  This is confirmed in numerical simulations of SNe~IIn (van
Marle et al.\ 2010).  When the density of pre-shock CSM fell rapidly,
the forward shock (which still contained $\sim$3$\times$10$^{49}$ ergs
of energy) would be free to expand.  The resulting speed is likely to
be highly latitude-dependent, of course.  Acceleration of the forward
shock in this way explains the absence of an obvious forward shock
immediately ahead of the Homunculus, and the existence of the very
rarefied cavity around the Homunculus at the present epoch.  This
acceleration of the forward shock essentially transfers the strong
thermal pressure support of the blast wave into kinetic energy for a
very small fraction of the total mass, thereby achieving speeds
significantly larger than the original maximum speed of ejecta in the
explosion.  For a few 10$^{49}$ ergs of energy in the forward shock,
rough estimates are $V$=5,000 km s$^{-1}$ and 0.1 $M_{\odot}$.

Although the hot material immediately behind the forward shock will
accelerate, most of the mass in the interaction was swept into the CDS
and would continue to coast at the same speed when CSM interaction
ends (the CDS does not accelerate because it is cold).  Despite being
compressed in a strong shock wave, the dense walls of the Homunculus
are seen today as neutral or molecular gas and cool dust.  This is
because their extremely high densities allowed for very rapid and
efficient cooling, which in turn powered the luminosity of the Great
Eruption.  The gas outside the Homunculus, however, is the result of a
strong shock accelerating through much lower-density gas outside the
pre-eruption CSM shell.  This material was heated by the shock and was
not able to cool efficiently, so it remains mostly ionized today.
Smith (2008) estimated an average density of $n_e \simeq $ 500
cm$^{-3}$ for the fast ejecta outside the Homunculus; for this
density, the recombination timescale

\begin{equation}
\tau \ = \ \frac{1}{n_e \ \alpha_B}  \Big{(} \frac{T}{10^4 K} \Big{)}^{0.8}
\end{equation}

\noindent (where $\alpha_B$=2.6$\times$10$^{-13}$ cm$^3$ s$^{-1}$ is
the hydrogen Case B recombination coefficient, and $T = 10^4$ K is the
ionized gas temperature) is about 250 yr.  Thus, the outer material
heated by the forward shock would still be ionized, whereas the much
denser material in the Homunculus would have long-since recombined.  A
strong shock passing through pre-existing dense clumps outside the
Homunculus may provide a possible explanation for the origin of the
long and thin whiskers/strings seen in {\it HST} images, but a
satisfying explanation requires detailed numerical hydrodynamic
simulations outside the scope of this paper. It is likely that
additional complications may be relevant in explaining the complex
material outside the Homunculus, such as a reflected (reverse) shock
resulting from the impact of the fast forward shock against the slower
ejecta from a previous eruption.

\subsection{What triggered the explosion?}

By explaining this long list of observable peculiarities of $\eta$ Car
with this single simple CSM-interaction model adapted from SNe IIn, we
distill the litany of many unsolved questions and inconsistencies down
to one central mystery: {\it Why did Eta Car suffer a 10$^{50}$ erg
  explosion in 1843/1844?}  This question remains unsolved, since a
$\sim$10$^{50}$ erg explosion was assumed in our model.  A number of
potential physical mechanisms have been discussed (see, e.g., Smith et
al.\ 2011 and references therein), including an explosion resulting
from explosive burning of fresh fuel mixed into a shell burning layer.
Perhaps this was triggered by a stellar collision, since the 1843
event coincided with a close periastron passage, although a collision
alone supplies insufficient power (Smith 2011; Smith et al.\ 2011).  A
great deal of theoretical work on stellar interiors is needed before
this is understood.  Adopting a single explosion does, however, make
the problem of $\eta$ Car more tractable than having a large number of
unrelated mysteries, if a physical explanation for such an explosion
can be identified.  Preliminary work does point to a tendency for very
massive stars to undergo hydrodynamic eruptions/explosions (Young et
al.\ 2005; Arnett, Meakin, \& Young 2005).

\section{IMPLICATIONS: EXTRAGALACTIC SUPERNOVA IMPOSTORS}

Since $\eta$~Car's Great Eruption occured in the mid-19th century
before modern optical spectrographs were available, and long before we
were able to measure IR or X-ray radiation, we are limited to
interpreting the historical visual light curve (Smith \& Frew 2011).
The fact that CSM interaction has long-since ended makes it difficult
to test some of the predictions of a CSM-interaction model directly,
which is why the Homunculus is such a valuable reservoir of physical
information.  This situation may soon change as we receive more
valuable information about the spectrum of $\eta$~Car's eruption
through spectroscopy of its light echoes (Rest et al.\ 2012).  In the
mean time, however, we can also consider implications of the
CSM-interaction scenario for transient sources that are thought to be
extragalactic analogs of $\eta$~Car.

Non-terminal eruptions or explosions that are analogous to $\eta$ Car
go by many names, including Type V supernovae, SN impostors, $\eta$
Car analogs, intermediate luminosity optical/red transients, eruptive
LBV-like transients, etc.\ (see Smith et al.\ 2011 and Van Dyk \&
Matheson 2012 for recent reviews).

If the CSM-interaction model is a viable interpretation of
$\eta$~Car's Great Eruption, it is likely that CSM interaction might
play a role in some other extragalactic LBV-like transients as well.
While some LBV-like transients exhibit the characteristic
F-supergiant-like spectrum that one expects from a dense
continuum-driven wind, a number of LBV-like eruptions do not fit that
bill.  There appear to be two classes, identified as ``hot LBVs''
exemplified by the SN impostor SN~2009ip, and ``cool LBVs''
exemplified by UGC2773-OT (see Smith et al.\ 2011, 2010b).  In
particular, SN~2009ip showed optical spectra that closely resemble
spectra of SNe~IIn, with evidence for fast-moving (1000-5000 km
s$^{-1}$) material seen only in absorption ahead of the photosphere
(Smith et al.\ 2010b; Foley et al.\ 2011).  This fast material might
correspond to the fast blast wave seen in $\eta$ Car's ejecta.  Like
$\eta$ Car, the light curve of SN~2009ip also showed a decade-long
increase in brightness leading up to a brief eruption peak of $-$14
mag (Smith et al.\ 2010), and it has since re-brightened twice (Drake
et al.\ 2010, 2012).  Perhaps the initial decade-long brightening of
SN~2009ip coincided with a strong wind that created a dense CSM, and
the brief brightening was the explosive initiation of a shock wave, as
in $\eta$ Car.  SN~2009ip is still under study, however.  In any case,
given the similarity between spectra of many LBV-like transients to
those of SNe~IIn, it is likely that CSM interaction may be quite
widespread.  Since the photosphere in many SNe~IIn is located ahead of
the forward shock in the very dense CSM, the same may be true for SN
impostors if CSM interaction is important.  If the emergent spectrum
is formed in the pre-shock wind in some cases, it may be possible to
determine the speeds of the pre-eruption wind and the expanding CDS
from the time evolution of spectra, as is done for SNe~IIn.
 
One important difference between $\eta$ Car and other LBV-like
transients is that most other SN impostors do not have high
luminosties that last for a decade or more; $\eta$ Car is quite
unusual in this respect.  In fact, most eruptive transients have
typical durations of only 100 days (Smith et al.\ 2011), like the 1838
and 1843 events of $\eta$ Car, but unlike its decade-long bright
phase.  We have argued that the long duration of high luminosity for
$\eta$~Car was caused by its explosion slowly overtaking very dense
CSM, produced in a strong continuum-driven wind that blew for the
preceding 30 years.  This long wind phase may be relatively rare,
since it is this phase that depends upon the star being extremely
massive and luminous, near the classical Eddington limit.  While
10$^{49}$-10$^{50}$ erg explosions may occur over a wide range of
initial masses (if, for example, the energy source is a deep-seated
explosive shell burning event), perhaps the preceding super-Eddington
wind phase - with sufficiently high mass-loss to make the CSM
interaction last for a decade - is limited to the most massive stars
like $\eta$~Car.  Lower-mass evolved stars may have dense and slow
winds that produce opaque CSM as well, but the range of radii over
which CSM densities are high-enough for efficient conversion of
kinetic energy to visual light may be small.  As a consequence,
non-terminal explosive transients from lower-mass stars might tend to
have shorter durations lacking prolonged CSM interaction, whereas the
longer-duration events might be limited to more massive stars.

Although CSM interaction may be important in many of these transients,
tracers of strong shocks like X-ray emission will not necessarily be
detectable.  In a SN~IIn, much of the X-ray luminosity generated by
the shock occurs in an optically thick region, and so the X-ray
luminosity is reprocessed and escapes ultimately as visual-wavelength
continuum radiation.  This is why SNe~IIn can be very luminous in the
visual continuum.  At late phases, when the forward shock outruns the
outer boundary of the CSM shell and accelerates, the X-rays may indeed
escape, but the X-ray luminosity may be too faint to detect in most
extragalactic transients.  Likely the most fruitful avenue for
searching for signs of CSM interaction is in detailed study of the
visual-wavelength spectra of LBV-like transients.

\section{SUMMARY}

This paper has examined the hypothesis that the Great Eruption of
$\eta$ Carinae was powered by CSM interaction, where ejecta kinetic
energy is converted to visual-wavelength luminosity in a radiative
shock passing through dense CSM.  By invoking an explosion of almost
10$^{50}$ ergs occuring in 1843 that expands into dense CSM, the
luminosity generated by CSM interaction provides an acceptable
explanation for the 10-15 yr bright phase of the Great Eruption.  We
found that a CSM which is created by a slow 200 km s$^{-1}$ wind with
$\dot{M}$=0.33 $M_{\odot}$ yr$^{-1}$ provides a much better match to
the historical light curve than CSM created by a previous explosion,
although the parameters of the simple model can be adjusted somewhat.
The wind needed to create the CSM is well-explained by existing models
of continuum-driven super-Eddington winds (Owocki et al.\ 2004; van
Marle et al.\ 2008, 2009), and the speed matches expectations for the
star's slower escape speed in the 1830s, based on its visual color and
magnitude at that time.  It also matches the observed wind speed seen
in spectra of the 1890 eruption.  Similarly, the physics of CSM
interaction to explain the observed luminosity is taken from standard
models for SNe~IIn.

Borrowing again from models and observations of SNe IIn, we find that
the CSM-interaction hypothesis can also explain a large number of
previously perplexing observed properties of the Homunculus nebula.
The most important consequence of CSM interaction is the creation of a
cold dense shell (CDS) where most of the mass resides, which we
identify with the thin walls of the massive Homunculus nebula.  We
find that the single scenario of CSM interaction gives a plausible
explanation for (1) the ratio of kinetic energy of the Homunculus to
the total radiated energy, (2) the double-shell structure of the
Homunculus, with most of the mass in a thin outer shell containing
cool dust and molecules, (3) the Hubble-like expansion of the
Homunculus, (4) the bipolar shape of the Homunculus, with the caveat
that published colliding-wind models need to be modified to limit the
CSM interaction to a decade after the explosion with more mass and a
highly radiative shock, (5) the efficient formation of dust grains in
a thin shell, analogous to the rapid dust formation observed in
SNe~IIn and SNe~Ibn, and (6) the acceleration of the forward shock
upon reaching the outer boundary of the dense CSM, which leads to the
very fast material in the outer ejecta and the soft X-ray shell around
$\eta$ Car.  Each of these requires a different explanation, or
mutually exclusive physical parameters in wind-only or explosion-only
scenario for $\eta$ Car, but they all arise naturally in the single
model of CSM interaction.

Lastly, we speculate that the phenomenon of CSM interaction might be
more widespread, and may play an important role in the physics and
observed properties of a number of non-terminal LBV-like transients
currently being discovered in external galaxies.  Many of these non-SN
transients may arise from weak (10$^{48}$ - 10$^{50}$ erg) explosions.
Indeed, some clear observational evidence supporting the presence of
strong CSM interaction is already seen in some LBV-like transients.
The presence of the same shock physics and observed consequences from
CSM interaction further blur the distinction between true
core-collapse SNe~IIn and their non-terminal impostors, changing the
framework in which we interpret LBV-like eruptions.  This may make it
more difficult to draw firm conclusions as to the underlying physical
nature of any observed event (i.e. core collapse or non-terminal)
based on the observed light curves and spectra, since both may be
dominated by radiation from CSM interaction.

\smallskip\smallskip\smallskip\smallskip
\noindent {\bf ACKNOWLEDGMENTS}
\smallskip
\footnotesize

I thank Stan Owocki for a number of thoughful conversations about
winds and LBV eruptions over the past decade, and for helpful comments
on the manuscript.  I also thank Dave Arnett for relevant discussions
and helpful comments on the paper.  Partial support was provided by the
National Aeronautics and Space Administration (NASA) through grant
AR-12618 from the Space Telescope Science Institute, which is operated
by AURA, Inc., under NASA contract NAS5-26555.  Based in part on
observations obtained at the Gemini Observatory, which is operated by
AURA, Inc., under a cooperative agreement with the NSF on behalf of
the Gemini partnership: the National Science Foundation (USA), the
Particle Physics and Astronomy Research Council (UK), the National
Research Council (Canada), CONICYT (Chile), the Australian Research
Council (Australia), CNPq (Brazil), and CONICET (Argentina).



\begin{thebibliography}
\scriptsize


\bibitem[]{} Aerts, C., Lamers, H.J.G.L.M., \& Molenberghs, G.\ 2004,
  A\&A, 418, 639

\bibitem[]{} Arnett, D., Meakin, C., \& Young, P.A.\ 2005, in ASP
Conf.\ Ser.\ 332, The Fate of the Most Massive Stars, ed.\ R.M.\
Humphreys \& K.Z.\ Stanek (San Francisco: ASP), 75

\bibitem[]{} Balbus, S.A., \& Schaan, E.\ 2012, preprint (arXiv:1207.3810)

\bibitem[]{} Belyanin, A.A.\ 1999, A\&A, 344, 199

\bibitem[]{} Castor, J.I., Abbott, D.C., \& Klein, R.I.\ 1975, ApJ,
195, 157

\bibitem[]{} Chevalier, R.A.\ 1992, ApJ, 394, 599

\bibitem[]{} Chevalier, R.A., \& Fransson, C.\ 1994, ApJ, 420, 268

\bibitem[]{} Chevalier, R.A., \& Irwin, C.M.\ 2011, ApJ, 729, L6

\bibitem[]{} Chugai, N.N., \& Danziger, I. J.\ 1994, MNRAS, 268, 173

\bibitem[]{} Chugai, N.N., et al.\ 2004, MNRAS, 352, 1213

\bibitem[]{} Clayton, D.D.\ 1979, Ap\&SS, 65, 179

\bibitem[]{} Colgate, S.A., \& Johnson, M.H.\ 1960, Phys.\ Rev.\ Lett., 5, 235

\bibitem[]{} Corcoran, M.F., Rawley, G.L., Swank, J.H., \& Petre, R.\
  1995, ApJ, 445, L121

\bibitem[]{} Corcoran, M.F., et al.\ 2004, ApJ, 613, 381

\bibitem[]{} Currie, D.G., et al.\ 1996, AJ, 112, 1115

\bibitem[]{} Crowther, P.A.\ 2007, ARAA, 45, 177

\bibitem[]{} Davidson, K.\ 1987, ApJ, 317, 760

\bibitem[]{} Davidson, K., \& Humphreys, R.M.\ 1997, ARAA, 35, 1

\bibitem[]{} Davidson, K., Walborn, N.R., \& Gull, T.R.\ 1982, ApJ, 254, L47

\bibitem[]{} Davidson, K., Dufour, R.J., Walborn, N.R., \& Gull, T.R.\
  et al. 1986, ApJ, 305, 867

\bibitem[]{} Dessart, L., Hillier, D.J., Gezari, S., Basa, S., \&
  Matheson, T.\ 2009, MNRAS, 394, 21

\bibitem[]{} Drake, A.J., et al.\ 2010, ATel, 2897, 1

\bibitem[]{} Drake, A.J., et al.\ 2012, ATel, 4334, 1

\bibitem[]{} Duschl, W.J., Hofman, K.H., Rigaut, F., \& Weigelt, G.\
  1995, in The Eta Carinae Region, ed.\ V.\ Niemala, N.\ Morrell, \&
  A.\ Feinstein (RevMexAA Ser. Conf., 2) (Mexico, D.F.: Inst.\
  Astron.\ Univ.\ Nac.\ Autonoma Mexico), 17

\bibitem[]{} Dwarkadas, V., \& Balick, B.\ 1998, AJ, 116, 829

\bibitem[]{} Dwarkadas, V., \& Owocki, S.P.\ 2002, ApJ, 581, 1337

\bibitem[]{} Falk, S.W., \& Arnett, W.D.\ 1977, ApJS, 33, 515

\bibitem[]{} Frank, A., \& Mellema, G.\ 1994, ApJ, 430, 800
 
\bibitem[]{} Frank, A., Balick, B., \& Davidson, K.\ 1995, ApJ, 441,
  L77
 
\bibitem[]{} Frank, A., Ryu, D., \& Davidson, K.\ 1998, ApJ, 520, 291

\bibitem[]{} Fryxell, B., M\"uller, E., \& Arnett, D.\ 1991, ApJ, 367,
  619

\bibitem[]{} Foley, R.J., et al. 2011, ApJ, 732, 32

\bibitem[]{} Fox, O.D.,et al.\ 2009, ApJ, 691, 650

\bibitem[]{} Fox, O.D., et al.\ 2011, ApJ, 741, 7

\bibitem[]{} Gallagher, J.S.\ 1989 in IAU Coll.\ 113: Physics of
  Luminous Blue Varibales, ed.\ K.\ Davidson, A.F.J.\ Moffat, \&
  H.J.G.L.M.\ Lamers (Dordrecht: Kluwer), 185

\bibitem[]{} Gandel'man, G.M., \& Frank-Kamenetsky, D.A.\ 1956, Sov.\
  Phys.\ Dokl., 1, 223

\bibitem[]{} Gomez, H.L., Dunne, L., Eales, S.A., \& Edmunds, M.G.\
  2006, MNRAS, 372, 1133

\bibitem[]{} Gomez, H.L., Vlahakis, C., Stretch, C.M., Dunne, L.,
  Eales, S.A., Beelen, A., Gomez, E.L., \& Edmunds, M.G.\ 2010, MNRAS,
  401, L48

\bibitem[]{} Gonzalez, R.F., de Gouveia Dal Pino, E.M., Raga, A.C., \&
  Velazqez, P.F.\ 2004, ApJ, 616, 976

\bibitem[]{} Gonzalez, R.F., Villa, A.M., Gomez, G.C., de Gouveia Dal
  Pino, E.M., Raga, A.C., Canto, J., Velazqez, P.F., \& de la Fuente,
  E.\ 2010, MNRAS, 402, 1141

\bibitem[]{} Gull, T.R., Viera, G., Bruhweiler, F., Nielsen, K.E.,
  Verner, E., \& Danks, A.\ 2005, ApJ, 620, 442

\bibitem[]{} Hartman, H., Gull, T., Johansson, S., Smith, N., and HST
  Eta Carinae Treasury Project Team 2004, A\&A, 419, 215

\bibitem[]{} Hillier, D.J., Davidson, K., Ishibashi, K., \& Gull,
T.R.\ 2001, ApJ, 553, 837

\bibitem[]{} Humphreys, R.M., Davidson, K., \& Smith, N.\ 1999, PASP,
  111, 1124

\bibitem[]{} Imshennik, V.S., \& Nad\"ezhin, D.K.\ 1989, Sov.\ Sci.\
  Rev.\ E.\ Astrophys.\ Space Phys., 8, 1

\bibitem[]{} Joss, P.C., Salpeter, E.E., \& Ostriker, J.P.\ 1973, ApJ,
181, 429

\bibitem[]{} Kankare, E., et al.\ 2012, MNRAS, 424, 855

\bibitem[]{} Kochanek, C.S.\ 2011, ApJ, 743, 73

\bibitem[]{} Kochanek, C.S.\ 2012, preprint (arXiv:1202.0281)

\bibitem[]{} Langer, N., Garcia-Segura, G., \& Mac Low, M.M.\ 1999,
  ApJ, 520, L49

\bibitem[]{} Madura, T.I., Gull, T.R., Owocki, S.P., Groh, J.H.,
  Okazaki, A.T., \& Russell, C.M.P.\ 2012, MNRAS, 420, 2064

\bibitem[]{} Matzner, C.D., \& McKee, C.F.\ 1999, ApJ, 510, 379

\bibitem[]{} Mauerhan, J., Smith, N., Silverman, J.M., Filippenko,
  A.V., Morgan, A. N., Cenko, S.B., Ganeshalingam, M., Clubb, K., \&
  Matheson, T.\ 2012, preprint (arXiv:1209.0821)

\bibitem[]{} Mauerhan, J., \& Smith, N.\ 2012, MNRAS, 424, 2659

\bibitem[]{} Meaburn, J., et al.\ 1996, MNRAS, 282, 1313

\bibitem[]{} Mellema, G., \& Frank, A.\ 1995, MNRAS, 273, 401

\bibitem[]{} Meshkov, E.F.\ 1969, Sov.\ Fluid Dyn., 4, 101

\bibitem[]{} Morse, J.A., Davidson, K., Bally, J., Ebbets, D., Balick,
  B., \& Frank, A.\ 1998, 116, 2443

\bibitem[]{} Morse, J.A., Kellogg, J.R., Bally, J., Davidson, K.,
Balick, B., \& Ebbets, D.\ 2001, ApJ, 548, L207

\bibitem[]{} Nielsen, K.E., Gull, T.R., \& Veira-Kober, G.\ 2005,
  ApJSS, 157, 138

\bibitem[]{} Ofek, E.O., et al.\ 2007, ApJ, 659, L13

\bibitem[]{} Ostriker, J.P., \& McKee, C.F.\ 1988, Rev.\ Mod.\ Phys.,
  60, 1

\bibitem[]{} Owocki, S.P.\ 2003, IAU Symp.\ 212, A Massive Star
  Odyssey, ed. K. van der Hucht, A. Herrero, \& C. Esteban (San
  Francisco: ASP), 281

\bibitem[]{} Owocki, S.P., Cranmer, S., \& Gayley, K.G.\ 1996, ApJ,
  472, L115

\bibitem[]{} Owocki, S.P., \& Gayley, K.G.\ 1997, in ASP Conf.\ Ser.\
  120, Luminous Blue Variables: Massive Stars in Transition, ed.\ A.\
  Nota \& H.\ Lamwers (San Francisco: ASP), 121

\bibitem[]{} Owocki, S.P., Gayley, K.G., \& Cranmer, S.\ 1998, in ASP
  Conf.\ Ser.\ 131, Boulder-Munich II: Properties of Hot, Luminous
  Stars, ed.\ I.D. Howarth (San Frncisco: ASP), 237

\bibitem[]{} Owocki, S.P., Gayley, K.G., \& Shaviv, N.J.\ 2004, ApJ,
616, 525

\bibitem[]{} Paczynski, B.\ 1990, ApJ, 363, 218

\bibitem[]{} Polomski, E., et al.\ 1999, AJ, 118, 2369

\bibitem[]{} Quinn, T., \& Paczynski, B.\ 1985, ApJ, 289, 634

\bibitem[]{} Rest, A., et al.\ 2012, Nature, 482, 375

\bibitem[]{} Richtmyer, R.D.\ 1960, Comm.\ on Pure and Appl. Math., 13, 297

\bibitem[]{} Roming, P.W.A., et al.\ 2012, ApJ, 751, 92

\bibitem[]{} Sakurai, A.\ 1960, Comm.\ Pure Appl.\ Math., 13, 353

\bibitem[]{} Sedov, L.I.\ 1959, Similarity and Dimensional Methods in
  Mechanics (New York: Academic)

\bibitem[]{} Seward, F.D., Butt, Y.M., Karovska, M., Prestwich, A.,
  Schlegel, E.M., \& Corcoran, M.F.\ 2001, ApJ, 553, 832

\bibitem[]{} Shaviv, N.J.\ 2000, ApJ, 532, L137


\bibitem[]{} Shacham, T., \& Shaviv, N.J.\ 2012, preprint
  (arXiv:1206.6078)

\bibitem[]{} Smith, N.\ 2002, MNRAS, 337, 1252


\bibitem[]{} Smith, N.\ 2006, ApJ, 644, 1151

\bibitem[]{} Smith, N.\ 2008, Nature, 455, 201

\bibitem[]{} Smith, N.\ 2009, arXiv:0906.2204

\bibitem[]{} Smith, N.\ 2010, MNRAS, 402, 145

\bibitem[]{} Smith, N.\ 2011, MNRAS, 415, 2020

\bibitem[]{} Smith, N., \& Gehrz, R.D.\ 1998, AJ, 116, 823

\bibitem[]{} Smith, N., \& Ferland, G.J.\ 2006, ApJ, 655, 911

\bibitem[]{} Smith, N., \& Frew, D. 2011, MNRAS, 415, 2009


\bibitem[]{} Smith, N., \& Mauerhan, J.C.\ 2012, ATel, 4412, 1

\bibitem[]{} Smith, N., \& McCray, R.\ 2007, ApJ, 671, L17

\bibitem[]{} Smith, N., \& Morse, J.A.\ 2004, ApJ, 605, 854

\bibitem[]{} Smith, N., \& Owocki, S.P.\ 2006, ApJ, 645, L45

\bibitem[]{} Smith, N., \& Townsend, R.H.D.\ 2007, ApJ, 666, 967

\bibitem[]{} Smith, N., Gehrz, R.D., \& Krautter, J.\ 1998, AJ, 116, 1332

\bibitem[]{} Smith, N., Gehrz, R.D., Hinz, P.M., Hoffmann, W.F.,
  Mamajek, E.E., \& Meyer, M.R., \& Hora, J.L.\ 2002, ApJ, 567, L77

\bibitem[]{} Smith, N., Gehrz, R.D., Hinz, P.M., Hoffmann, W.F., Hora,
J.L., Mamajek, E.E., \& Meyer, M.R.\ 2003b, AJ, 125, 1458


\bibitem[]{} Smith, N., et al.\ 2004, ApJ, 605, 405

\bibitem[]{} Smith, N., Morse, J.A., \& Bally, J.\ 2005, AJ, 130, 1778


\bibitem[]{} Smith, N., et al.\ 2007, ApJ, 666, 1116 

\bibitem[]{} Smith, N., et al.\ 2008a, ApJ, 686, 467 

\bibitem[]{} Smith, N., Foley, R.J., \& Filippenko, A.V.\ 2008b, ApJ,
  680, 568 

\bibitem[]{} Smith, N., et al.\ 2009, ApJ, 695, 1334 

\bibitem[]{} Smith, N., Chornock, R., Silverman, J.M., Filippenko,
  A.V., \& Foley, R.J.\ 2010, ApJ, 709, 856 

\bibitem[]{} Smith, N., et al.\ 2010, AJ, 139, 1451 

\bibitem[]{} Smith, N., Li, W., Silverman, J.M., Ganeshalingam, M., \&
  Filippenko, A.V.\ 2011, MNRAS, 415, 773 

\bibitem[]{} Smith, N., et al.\ 2012, AJ, 143, 17 

\bibitem[]{} Soker, N.\ 2007, ApJ, 661, 490

\bibitem[]{} Van Dyk, S.D.\ 2005, in ASP Conf.\ Ser.\ 332, The Fate of
the Most Massive Stars, ed.\ R.M.\ Humphreys \& K.Z.\ Stanek (San
Francisco: ASP), 47

\bibitem[]{} Van Dyk, S.D., \& Matheson, T.\ 2012, in Eta Carinae and
  the Supernova Impostors, ASSL 384, 249

\bibitem[]{} van Marle, A.J., Owocki, S.P., Shaviv, N.J.\ 2008, MNRAS,
  389, 1353

\bibitem[]{} van Marle, A.J., Owocki, S.P., Shaviv, N.J.\ 2009, MNRAS, 394, 595

\bibitem[]{} van Marle, A.J., Smith, N., Owocki, S.P., \& van Veelen,
  B.\ 2010, MNRAS, 407, 2305

\bibitem[]{} Vishniac, E.T.\ 1994, ApJ, 428, 186

\bibitem[]{} Walborn, N.R.\ 1976, ApJ, 204, L17

\bibitem[]{} Walborn, N.R., \& Liller, M.\ 1977, ApJ, 211, 181

\bibitem[]{} Walborn, N.R., \& Blanco, B.M.\ 1988, PASP, 100, 797

\bibitem[]{} Walborn, N.R., Blanco, B.M., \& Thackery 1978, ApJ, 219, 498

\bibitem[]{} Weis, K., Duschl, W.J., \& Chu, Y.H.\ 1999, A\&A, 349, 467

\bibitem[]{} Weis, K.\ 2001, in ASP Conf.\ Ser.\ 242, Eta Carinae and
  other Mysterious Stars, ed.\ T.\ Gull, S.\ Johansson, \& K.\
  Davidson (San Francisco: ASP), 129

\bibitem[]{} Weis, K., Corcoran, M.F., Bomans, D.J., \& Davidson, K.\
  2004, A\&A, 415, 595

\bibitem[]{} Whitney, C.A.\ 1952, Harvard Obs.\ Bull., 921, 8

\bibitem[]{} Woosley, S.P., Blinnekov, S., \& Heger, A.\ 2007, Nature, 450, 390

\bibitem[]{} Young, P.A.\ 2005, in ASP Conf.\ Ser.\ 332, The Fate of
the Most Massive Stars, ed.\ R.M.\ Humphreys \& K.Z.\ Stanek (San
Francisco: ASP), 190

\bibitem[]{} Zethson, T., Johansson, S., Davidson, K., Humphreys,
  R.M., Ishibashi, K., \& Ebbets, D.\ 1999, A\&A, 344, 211

\bibitem[]{} Zwicky, F.\ 1964, ApJ, 139, 514

\end{thebibliography}
\end{document}